\documentclass{aa}
\usepackage{graphics,supertabular}
\usepackage{longtable}
\usepackage{graphicx}
\usepackage{amsfonts}
\usepackage{natbib}
\usepackage{amssymb}
\usepackage{hyperref}
\usepackage{subfigure}
\usepackage[font=small]{caption}
\usepackage{csvsimple}
\usepackage{booktabs}
\usepackage{pgfkeys} 
\usepackage{algorithm2e}
\usepackage{xcolor}
\usepackage{caption}
\usepackage{csvsimple}
\usepackage{booktabs}
\usepackage{pgfkeys} 
\usepackage{algorithm2e}
\usepackage{xcolor}

\begin{document}

\title{Multiwavelength study of the Carina--Sagittarius Arm}
\subtitle{I. Astrometric and photometric search for new open clusters in the $320^{\circ} \leq l \leq 325^{\circ}$ region}

\author{M.A. Corti\inst{1,2}\thanks{Member of the {\bf ES}tructuras de HI y {\bf F}ormaci\'on {\bf E}stela{\bf R} g{\bf A}l\'actica (ESFERA) group, Argentina.}, L. Rizzo\inst{1}\protect\footnotemark[1], L.G. Pa\'iz\inst{1,2}\protect\footnotemark[1] \& G.S. Elizalde Caviglia\inst{3}\protect\footnotemark[1]}

\authorrunning{Corti et al.}
\titlerunning{New open clusters in the Carina-Sagittarius Arm}
%\contact{ACR: mariela@fcaglp.unlp.edu.ar}

\institute{
Facultad de Ciencias Astron\'omicas y Geof\'{\i}sicas,
(UNLP), La Plata, Argentina
\and 
Instituto de Astrof\'{\i}sica La Plata, (CONICET - UNLP), La Plata, Argentina
\and
Facultad Regional Concepci\'on del Uruguay, (UTN), Entre R\'ios, Argentina}

\date{Received \today; accepted \ldots }
\abstract
{}
{Our main goal is to determine whether the lower stellar density observed in the Carina–Sagittarius Arm region ($320^{\circ} \leq l \leq 325^{\circ}$, $|b| \leq 1^{\circ}$) is an intrinsic structural feature of the spiral arm or a consequence of high interstellar reddening and extinction along the line of sight.}
{We performed a systematic search for new open clusters using the HDBSCAN algorithm on \textit{Gaia} DR3 astrometric data. The physical reality of the candidates was validated through a multiwavelength analysis, integrating \textit{Gaia} photometry with NIR of the 2MASS catalog, MIR photometric information of the WISE catalog to identify young stellar objects (YSOs), and the radio continuum emission at 843 MHz with SUMSS survey to identify and associated interstellar material.}
{We report the discovery of five new open clusters, the \textit{ESFERA} sample. Our analysis reveals four young systems (10–50~Myr), with \textit{ESFERA\,3} hosting YSOs, while \textit{ESFERA\,5} is identified as a 2~Gyr-old cluster. Although the region lacks massive star-forming complexes, these clusters trace active, despite the fact that lower-efficiency, star formation within the region.}
{Our results suggest that this region represents an intrinsic star-formation "valley" or low-density node (the "string" in a "beads on a string" morphology) rather than an effect of extinction. The discovery of these clusters demonstrates that while the local surface density is below the critical threshold for massive complexes, star formation persists in isolated, small-scale pockets throughout the arm. }


\maketitle
%
%_______________________________________________________________

\section{Introduction}

Young open clusters (OCs) constitute primary tracers of the Milky Way's spiral structure, offering a detailed sampling of the star formation history in the Galactic disk. As intrinsically bright systems that host massive, early-type stars (O and B), they offer a distinct advantage for Galactic studies, as their young ages ensure they remain close to their birthplaces in the spiral arms. Furthermore, their abundance throughout the disk provides a valuable and widely distributed sample for tracing the Galactic disc's properties, complementing other tracers such as Cepheids and masers~\citep{efremov1998, 2008va, 2014choi, Molina2018, 2021dias}.

Crucially, distances to open clusters can be determined with significantly higher precision than those of individual stars. By leveraging both the photometry and kinematics of multiple cluster members, it is possible to mitigate individual measurement uncertainties. This collective analysis provides a more accurate distance estimate based on the statistical consensus of all member stars. Recent studies have highlighted an intriguing anomaly in the spatial distribution of young stellar tracers. Specifically, large-scale surveys of OCs and OB stars \citep[e.g.,][]{2019Chen, 2025liu} have revealed a noticeable `gap' or scarcity of detections within the Carina–Sagittarius Arm. This persistent lack of tracers at certain Galactic coordinates and the distance ranges raises fundamental questions regarding the continuity and local structure of this spiral arm, suggesting that our current census of the region remains incomplete.

This work aims to bridge this observational gap by conducting a systematic search for new open clusters within the Carina–Sagittarius Arm, specifically targeting the region $320^{\circ} \leq l \leq 325^{\circ}$. While the number of cataloged OCs has grown significantly with the advent of the \textit{Gaia} mission \citep{Gaia2016}, many candidates lack robust astrometric and physical confirmation. A major challenge in current Galactic studies is the presence of spurious entries, including duplicate records and chance alignments of field stars (asterisms) that do not constitute true clusters. Such inaccuracies can bias our understanding of the spiral structure. Consequently, we apply a rigorous membership analysis to identify genuine clusters, ensuring a reliable census for the local structure of this arm. 

The initial validation of open cluster candidates is as critical as the subsequent astrophysical analysis. To ensure maximum reliability, we adopt a multi-wavelength approach that extends beyond optical observations. This involves high-precision astrometric analysis using \textit{Gaia} Data Release 3 (\textit{Gaia} DR3) data --specifically parallaxes and proper motions-- to confirm the common distance and kinematic coherence of the candidate stars. Furthermore, we integrate optical photometry with near- and mid-infrared (NIR/MIR) data to mitigate the effects of interstellar extinction and derive robust stellar parameters, such as age and distance. This comprehensive approach allows for a rigorous cross-validation of each stellar aggregate, ensuring that only physically genuine open clusters are used to map the Carina–Sagittarius Arm. 

The paper is structured as follows. In Sect.\,2, we describe the observational data and the various catalogs and services used to obtain the astrometric and astrophysical parameters. Section\,3 details the candidate selection process, including the astrometric and astrophysical constraints applied. The analysis and results for each identified open cluster are presented in Sect.\,4. Finally, the discussion and conclusions are provided in Sects.\,5 and 6, respectively.

\section{Observational Data}
\label{astrom data}
\subsection{Astrometric Data}
We use the astrometric and photometric parameters from the {\it Gaia}\,DR3 astrometric catalog \citep{Gaia2023}. 
Released on 13 June 2022, 
it consists of more than $1.81\times10^{9}$ sources with a limiting magnitude of G $\sim 21$ and a bright limit of G $\sim 3$. It contains some 585 million sources with five-parameter astrometry: position $(\alpha,\ \delta)$, proper motion ($\mu_{\alpha}\ \rm{cos}\ \delta$,\ $\mu_{\delta}$) and parallax $\varpi$. It has about 882 million sources with six-parameter astrometry, including an additional pseudocolour parameter. The astrometric solution is accompanied with some quality indicators, like the Renormalised Unit Weight Error (RUWE). 

\subsection{Photometric data}
\label{photometric data}
The comprehensive study of open clusters necessitates both astrometric analysis to identify cluster members and estimate distances using parallax data and, photometric analysis across a wide range of wavelengths. Consequently, in addition to the astrometric data of each stellar member, we utilized the optical photometric data (G, $G_{\text{BP}}$, and $G_{\text{RP}}$ bands) provided  by {\it Gaia} DR3. The magnitudes are given for more than $1.80\times10^{9}$ in the G-band (330-1050 nm), for more than $1.54\times10^{9}$ sources in the G$_{BP}$-band (330-680 nm), and for more than $1.55\times10^{9}$ sources in the G$_{RP}$-band (640-1050 nm). The photometric uncertainties, supplied by the catalog, are conservatively constrained to $\sigma_G \leq$ 0.01 mag and, $\sigma_{(BP-RP)} \leq$ 0.1 mag. The Color-Magnitude Diagram (CMD) serves as the primary tool for estimating the essential photometric characteristics of a star cluster. 
For our analysis with the {\it Gaia} data, we employed the Padova stellar isochrones from the PARSEC v1.2S models\footnote{http://stev.oapd.inaf.it/cgi-bin/cmd}. These models, based on the stellar evolutionary tracks by Bressan et al. (2012), were computed for a scaled-solar composition following the {\rm Y} = 0.2485+1.78{\rm Z} relation, adopting a solar metallicity of Z = 0.0152. 

To investigate the reddened stars and Young Stellar Object (YSO) candidates within each cluster, we incorporated near$-$infrared (NIR) photometry ($JHK$ bands)  from the Two Micron All-Sky Survey (2MASS)\footnote{http://www.ipac.caltech.edu/2mass/} \citep{Skrutskie06}. The photometric uncertainties provided by the 2MASS catalogue are typically $\sigma_{(JHK)} \leq$ 0.05 mag. Additionally, we included mid-infrared (MIR) photometric data from the Wide-field Infrared Survey Explorer (WISE) catalog\footnote{http://irsa.ipac.caltech.edu/Missions/wise.html} \citep{Wri10}. This MIR information covers the 3.4, 4.6, 12, and 22 $\mu$m bands ($W_1$, $W_2$, $W_3$, and $W_4$, respectively).

Knowledge of the interstellar medium (ISM) is vital when studying a stellar formation region. To investigate the potential existence of an  H\,II region and/or an H\,I shell and their connection with the star clusters, we utilized data from the Sydney University Molonglo Sky Survey (SUMSS) \citep{boc99}. SUMSS provides radio continuum emission data at 843\,MHz. The survey's synthesized elliptical beam is $45'' \csc |\delta| \times 45''$. This information was specifically used to search for interactions between the stellar population and the surrounding gas structures.

To manage and analyze the astronomical data, we utilized several standard software tools:
\begin{itemize}
\item Data Management and Analysis: We employed the Virtual Observatory software TOPCAT (Tool for Operations on Catalogs and Tables)\citep{Taylor11} for efficient management and analysis of the astronomical tables.
\item Image Visualization: To analyze the images, we primarily used two visualization tools:
\begin{itemize}
\item KVIS \citep{Gooch96}, which is part of the KARMA (A Virtual Reality Toolkit for Astronomers) astronomical visualization toolkit\footnote{https://www.atnf.csiro.au/computing/software/karma/}.
\item SAOImage DS9, developed and maintained by the Smithsonian Astrophysical Observatory (SAO) as an essential component of its astronomical visualization toolkit \citep{Joye25}. 
\end{itemize}
\item Catalog Access: To search and access a comprehensive collection of published astronomical data catalogs across various wavelengths, we consulted the VizieR tool\footnote {https://vizier.cds.unistra.fr}.
\end{itemize}

\section{Data sources and membership}

\begin{figure}[!t] 
   \centering
\includegraphics[width=\columnwidth]{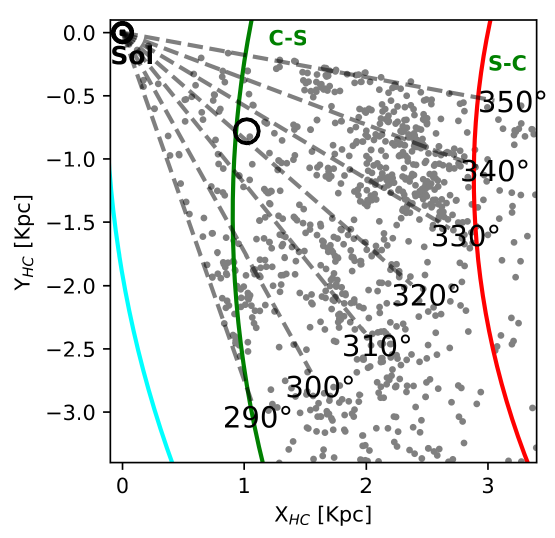}
\caption{Spatial distribution in heliocentric Galactic coordinates ($X_{HC}, Y_{HC}$) of known open clusters (gray points) from the UCC catalog \citep[]{Perren23}. The plot covers the region $290^\circ < l < 350^\circ$ and $-1^\circ < b < 1^\circ$, corresponding to a prominent section of the Carina-Sagittarius Arm. The colored curved lines represent the \citet{HouHan14} spiral structure of the MW. The thick black circle highlights the specific region investigated in this study, centered at $l=322^\circ$, which is characterized by a noticeable lack of known clusters and limited prior deep exploration.}

\label{fig:distribOCs}
    \end{figure}

\begin{figure}[!t] 
   \centering
\includegraphics[width=\columnwidth]{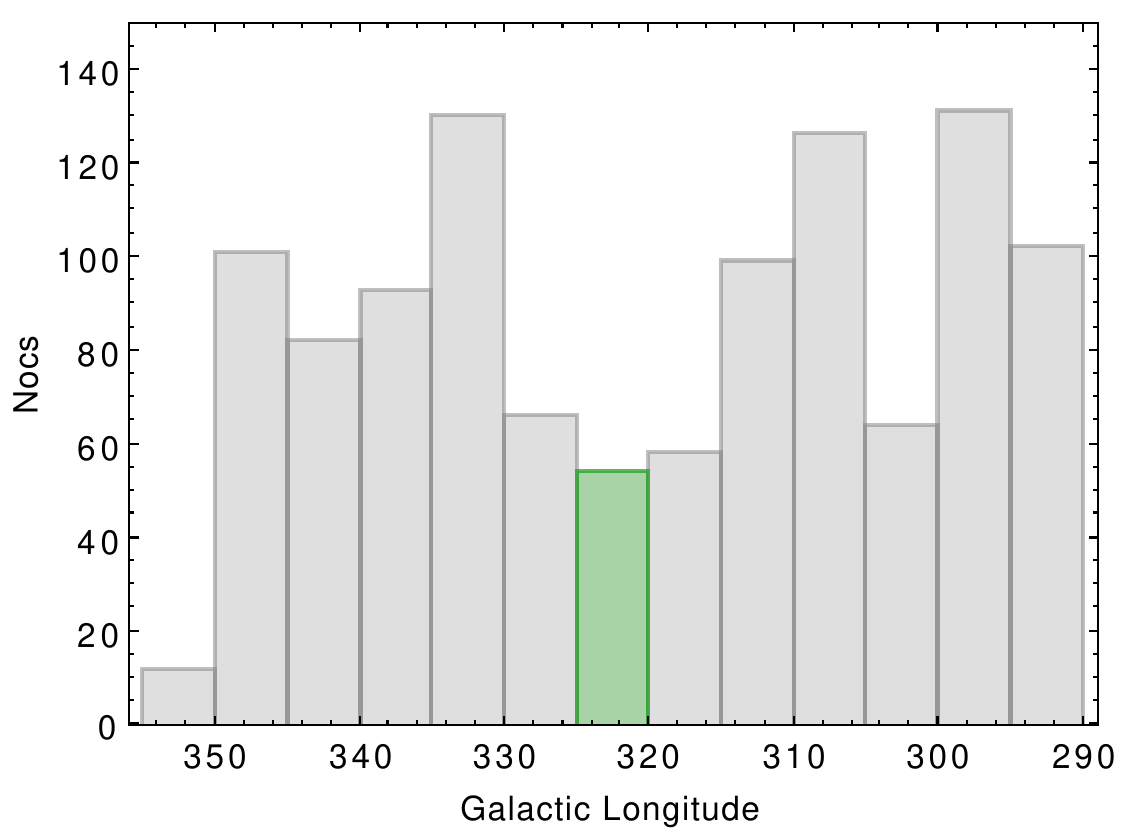}
\caption{Number of open clusters per $5^{\circ}$ interval in galactic longitude. This distribution highlights the density variations within the section of the Carina-Sagittarius Arm.}
\label{fig:HistogOCs}
   \end{figure}

We search for open clusters in the Carina-Sagittarius Arm region, more precisely $320^{\circ} \leq l \leq 325^{\circ}$, $-1^{\circ} \leq b \leq +1^{\circ}$. Figure~\ref{fig:distribOCs} shows the distribution of open clusters (from the UCC catalog \citet{Perren23}) in a section of the Carina-Sagittarius Arm . Figure~\ref{fig:HistogOCs} presents a histogram of the OCs shown in Fig.~\ref{fig:distribOCs} to provide a quantitative view of the region selected for this study. The data displays a clear lack of open clusters; this anomaly makes it a particularly compelling sector to investigate.

We use {\it Gaia} DR3 astrometric data (positions, proper motions and parallaxes) for sources with standard error in proper motion $\leq$ 0.3 mas yr$^{-1}$ to minimize the errors arising from this parameter \citep{2025esfera}; the quotient parallax divided by its standard error must be 
$\geq$ 5 (\texttt{parallax\_over\_error} $\geq 5$) to reduce the errors in the transformation of mean weighted parallax to distance \citep{Paiz2025};
and the local RUWE threshold was determined using the GaiaUnlimited python package \citep{2024cas} to identify and discard potential binary systems and sources with unreliable astrometric solutions. The final sample consists of 247296 sources. These stringent astrometric quality cuts naturally limit the sample to stars typically brighter than $G \approx 18$ mag, ensuring that the membership analysis and subsequent isochrone fitting are based on high-precision data. Although these criteria exclude the lower Main Sequence, the sampled population (Turn-Off and RGB) remains sufficient for a robust characterization of the clusters' physical parameters.

\subsection{Candidates Selection}
\label{candidates}
We normalize the five astrometric data ($\alpha$, $\delta$, $\mu_{\alpha}\ \rm{cos}\ \delta$, $\mu_{\delta}$, $\varpi$) through \texttt{RobustScaler} on the \texttt{scikit-learn} package \citep{pedregosa2011}; for example, it removes the median and can scale the data according to the interquartile range. After this, membership selection is performed using the HDBSCAN clustering algorithm \citep{campello2013} on the five normalized astrometric parameters using its hyperparameters \texttt{min\_samples} and \texttt{min\_cluster\_size} set to the same value (as recommended by \cite{campello2013}). For this, we explore the range 5-100 to ensure cluster stability/discovery; the selection method (\texttt{cluster\_selection\_method}) can be Excess of Mass (\texttt{eom}) to select one or two of the largest clusters and some smaller clusters, or \texttt{leaf} to select several small and more homogeneous clusters \citep{SantosSilva2021}. We chose \texttt{leaf} to prioritize the detection of small, homogeneous, and dense groupings; and with \texttt{metric=euclidean} we use the standard euclidean metric to measure the five-dimensional distance between points.

We consider the astrometric members of a group are those stars with HDBSCAN probability greater than or equal to 0.5. Once HDBSCAN finds groupings, we have to evaluate if these correspond to real clusters or false positives. We consider them as false positives if they do not fulfill some of the following conditions by \cite{Cantat2020Anders} and \cite{Hunt2021}:

a) the dispersion of the total proper motion of the group members must satisfy  

\begin{equation}
\label{dispersion}
\sqrt{\sigma_{\mu_{\alpha}cos{\delta}}^{2}+\sigma_{\mu_{\delta}}^{2}} \le \left\lbrace \begin{array}{l} 1 \hspace{0.1cm} \text{mas} \hspace{0.1cm} \text{yr}^{-1} \hspace{1.2cm} \text{if } \varpi \le  0.67 \hspace{0.1cm} \text{mas}\\ 1.49\cdot\varpi \hspace{0.1cm} \text{mas} \hspace{0.1cm} \text{yr}^{-1} \hspace{0.2cm} \text{if } \varpi >  0.67 \hspace{0.1cm} \text{mas}; \end{array} \right.
\end{equation}

b) the radius containing half of the group members $(r_{50})$ must be less than 20\,pc;
 
c) the group must be denser than a surrounding set of 100 to 500 field stars. To compare densities in the five-dimensional space of normalized astrometric data, we compute the euclidean distance from each star to its ninth nearest neighbor in the group. This is because we assume that open clusters consist of at least ten members, and the ninth neighbor is free of contamination from binary or multiple star systems.

\begin{figure*} 
   \centering
\includegraphics[width=0.7\columnwidth,angle=-90]{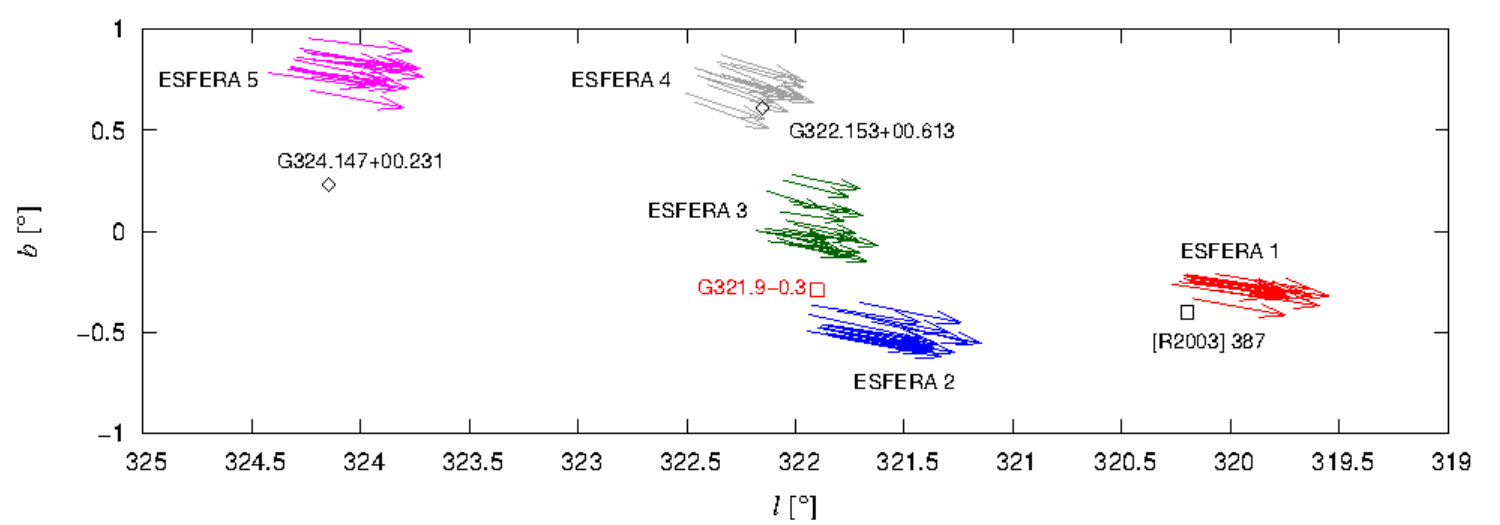}
\caption{Spatial distribution of the \textit{ESFERA} open cluster members with their proper motions in galactic coordinates. Different symbols represent HII region [R2003] 387\citep{Murray10} (black open square); G\,321.9$-$0.3 SNR \citep{Ball25} (red open square); HII regions G\,322.153$+$00.613 \citep{Anderson14} and G\,324.147+00.231 \citep{Anderson14} (black diamonds). Proper motion components in galactic coordinates were calculated from the formulae in \cite{Poleski2013}. The background is intentionally kept clear to maximize the contrast and visibility of the proper motion vectors.}
\label{fig:mulmub}
    \end{figure*}

In this way, we divide the region of study into five sub-regions of two square degrees. Out of the hundreds of groups initially identified by HDBSCAN across all sub-regions, 26 passed the astrometric filters (conditions a, b, and c), and 5 of them were confirmed as new open clusters with the isochrone fitting. This means we detect a total amount of five groupings representing new open clusters that we call \textit{ESFERA\,1}, \textit{ESFERA\,2}, \textit{ESFERA\,3}, \textit{ESFERA\,4}, and \textit{ESFERA\,5}. 
Fig.~\ref{fig:mulmub} shows the spatial distribution of the \textit{ESFERA} open cluster members with their proper motions in galactic coordinates calculated from the formulae in \cite{Poleski2013}.

\subsection{Physical Parameters Estimation}
\label{physical}
Building upon the sample of cluster member stars identified via the astrometric analysis described in Sect. \ref{candidates}, and adopting the cluster distances derived from their respective parallaxes, we initiated the photometric analysis for each open cluster by studying their CMDs using data from the {\it Gaia}\,DR3 survey.

To determine the visual absorption for each cluster, we utilized the theoretical Schmidt-Kaler Principal Sequence \citep{LB82} in conjunction with coefficients derived from the polynomial fittings in the color-color diagram, using the Johnson-Cousins passbands \citep{Jordi10}. These fittings are described by Eqs.\ref{colour} and \ref{magnitude}. The standard deviation of the residuals of the fit was adopted as the $\sigma$ value. The described theoretical curve was fitted to the {\it Gaia} photometric observations of the possible cluster members using $A_v$ values constrained by the range proposed by \citet{Joshi25}(see their Fig.4), for $300^{\circ} \leq l \leq 350^{\circ}$. 
\begin{equation}
\label{colour}
\begin{split}
    G_{BP} - G_{RP} = 0.0981 + 1.429 (B-V) - 0.0269 (B-V)^2 +\\
    0.0061(B-V)^3; \sigma = 0.43  
\end{split}
\end{equation}
\begin{equation}
\label{magnitude}
\begin{split}
G = V - 0.0424 - 0.0851(B-V) - 0.3348(B-V)^2 + \\
0.0205(B-V)^3; \sigma = 0.38
\end{split}
\end{equation}

Once the $A_v$ value was estimated for each open cluster, in order to determine their respective ages, the PARSEC v1.2S isochrones were fitted to their $G$ vs $G_{\text{BP}}$-$G_{\text{RP}}$ photometric diagrams. For this purpose, the reddening law ${A_G}$ = 0.83627\,${Av}$, ${A_{BP}}$ = 1.08337\,${Av}$, and ${A_{RP}}$ = 0.63439\,${Av}$, was applied. Since age estimates derived from broadband photometric surveys and small member counts are inherently uncertain, we fit the cluster prioritizing the alignment of the most probable cluster members with the Zero Age Main Sequence (ZAMS) and the identified early-type stars. Crucially, the metallicity was constrained to solar ($Z=Z_{\odot}$) across all these fitting attempts, as previously discussed in Sec.\ref{photometric data}. As one stellar formation region clearly exhibits differential reddening, we needed a method to distinguish between reddening caused by dust and a star's inherent color. Our protocol was to cross-reference the optical {\it Gaia} data (for stars confirmed as cluster members) with NIR photometric data from the 2MASS survey ($JHK$). We utilized a strict maximum angular distance of $1''$ for this matching process.
Subsequently, to avoid any confusion arising from the intrinsic degeneracy between reddening and spectral type, we calculate the reddening$-$free photometric parameter Q$_{NIR} = (J-H) -$ 1.7 $\times (H-K)$ \cite{Negueruela07}. Our star classification is substantiated by the guidelines provided by \cite{Borissova12} and \cite{Messineo12} where $-$0.3 $<$ Q$_{NIR} < $ 0.5 values are indicative of early main sequence stars, Q$_{NIR} < -$0.3 values are indicative of Pre-Main Sequence Stars (PMS) or Young Stellar Objects (YSO) and 0.5 $<$ Q$_{NIR}$ values are indicative of late reddening stars. Following this classification, potential YSO candidates ($Q_{NIR} < -0.3$) were further investigated by cross-matching them with the WISE catalog. This was done with the aim of confirming their YSO nature and subsequently classifying them according to the criteria established by \cite{Koenig14}. The interstellar extinction vector, derived from the reddening law defined by \citet{Wang19}, was superimposed on each CMD.
This vector represents the reddening correction required to align the observed data with the intrinsic (de-reddened) position of the theoretical isochrone, especially for the hotter and more luminous stellar populations. Based on this analysis, we estimated the spectral types of the cluster members and performed a classification using four categories: O$^-$, O$^+$, B$^-$, and B$^+$. In this nomenclature, the letters O and B denote the corresponding spectral types, while the minus ($-$) and plus ($+$) signs indicate early-type and late-type stars within each class, respectively.

To complete the study about the redder stars, we analyze the $(J-H)$ vs. $(H-K)$ color-color diagram to distinguish different stellar populations. In our procedure, we adopted the Main Sequence (MS) calibrations given by \cite{LB82}, \cite{Dean78}, and \cite{Koornneef83}. Their locations were computed using the adopted distances and visual absorption presented in Table~\ref{table:generalparam} and a normal reddening law, which allowed us to use the absorption ratios ($r_X = A_X/A_V$) given by \cite{Rieke85} and \cite{Carpenter01}. Candidate members for each cluster are ordered by increasing {\it Gaia} Source ID in all tables. This provides us a uniform reference across both, astrometric and photometric datasets, ensuring that specific stars can be easily identified and cross-referenced throughout the study.

Finally, the flux density at 843 MHz was estimated for each source located in the vicinity of each \textit{ESFERA} open cluster. Using the TVSTAT task within the AIPS\footnote{http://www.aips.nrao.edu/index.shtml} package, the average background flux density ($S_{bg}$) was determined in units of Jy \text{beam}$^{-1}$. This was achieved by calculating the mean flux density value at three different positions adjacent to each source. The root mean square (rms) of $S_{bg}$ was then used as the detection threshold to distinguish the signal from the noise. Subsequently, the measured flux density value of each source was obtained, and the corresponding $S_{bg}$ value was subtracted to yield the net emission. The spectral index $\alpha$ ($S \sim \nu^{-\alpha}$) is a key parameter that indicates the origin of the emission (either thermal radiation or non-thermal synchrotron radiation). It can be calculated from the flux densities measured at two different frequencies, $\nu_1$ and $\nu_2$ with the mathematic expression $\alpha = \frac{\log(S_1 / S_2)}{\log({\nu_2}/{\nu_1})}$. The determination of this spectral index is beyond the scope of this paper and will be addressed in a future study.

\section{Analysis and Results}
\label{analysis}

\begin{figure*}
   \centering
   \includegraphics[width=0.5\columnwidth,angle=-90]{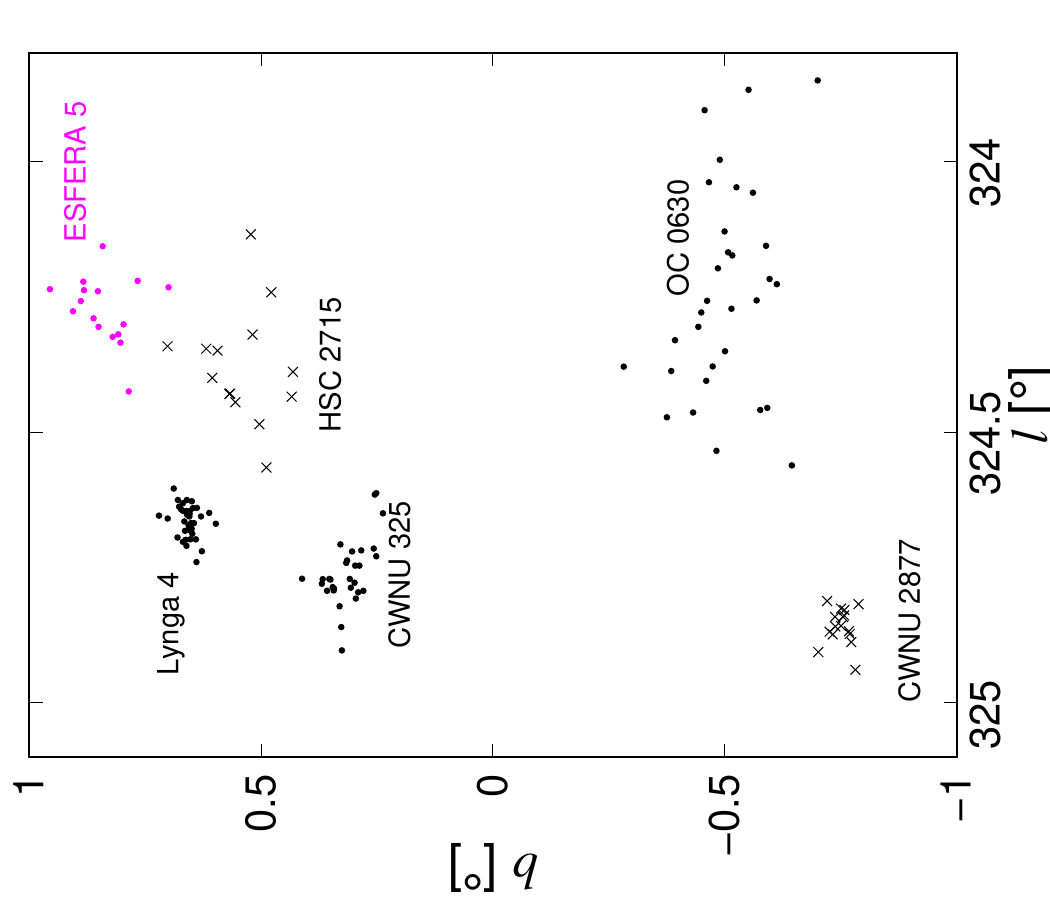}
   \includegraphics[width=0.5\columnwidth,angle=-90]{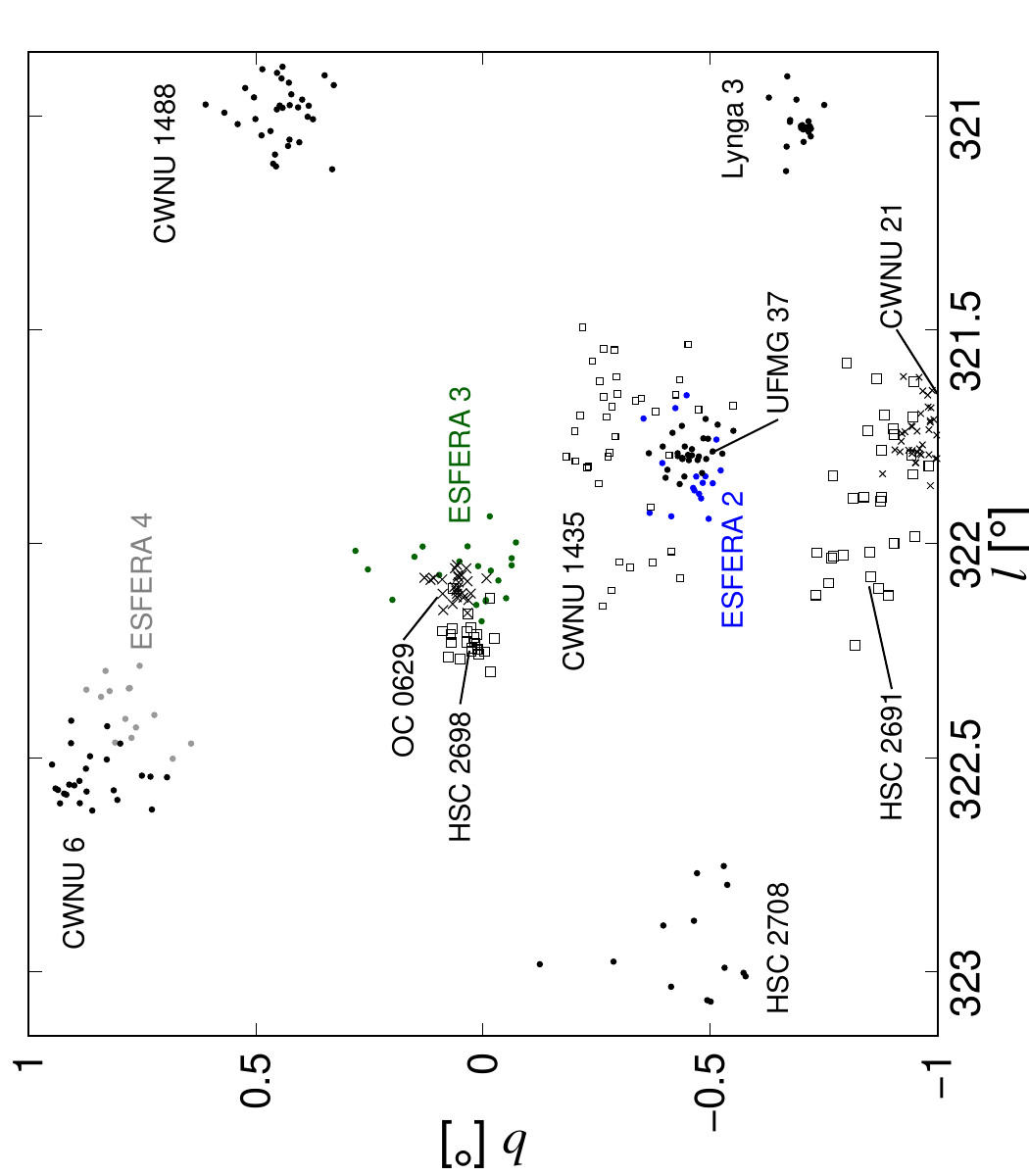}
   \includegraphics[width=0.5\columnwidth,angle=-90]{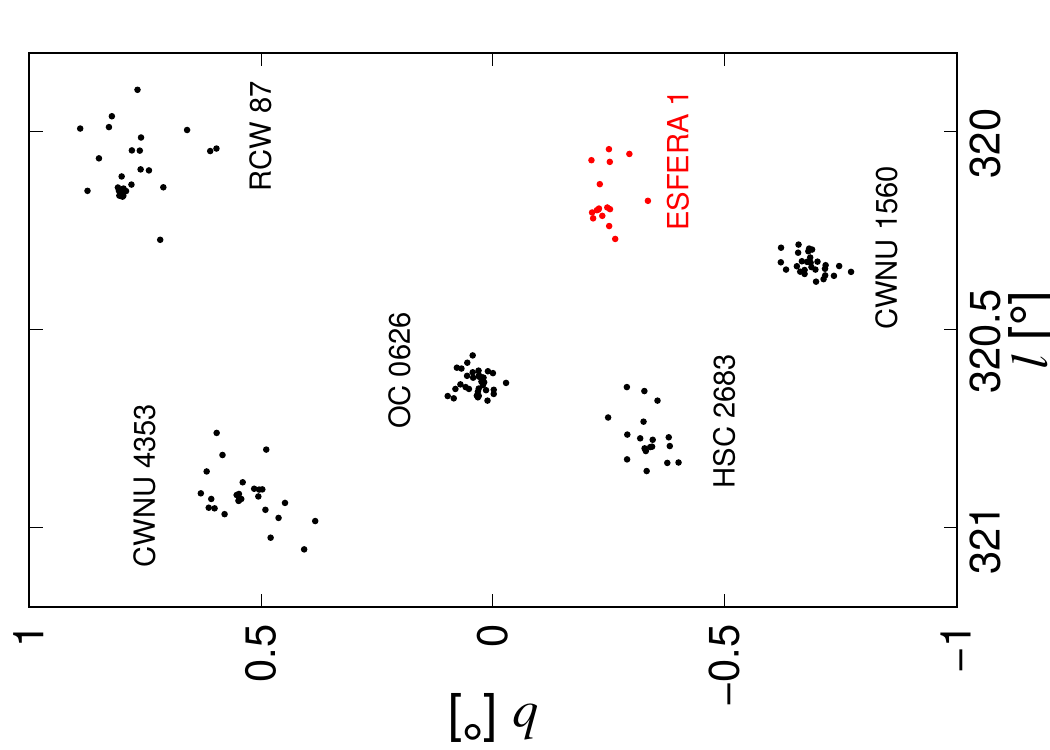}
\caption{Spatial distribution of \textit{ESFERA\,1} (red), \textit{ESFERA\,2} (blue), \textit{ESFERA\,3} (green), \textit{ESFERA\,4} (gray) and \textit{ESFERA\,5} (magenta) and local open clusters known in the literature (black) detected in this study.}
\label{fig:zona-Esf12345}
    \end{figure*}

In this Sect., we present the detection and characterization of five new open cluster candidates, hereafter designated as the \textit{ESFERA} sample. Our analysis successfully identified these stellar aggregates alongside several previously catalog clusters, confirming the robustness of our selection criteria. Figure~\ref{fig:zona-Esf12345} illustrates the spatial distribution of these new candidates in relation to known open clusters in the literature (\cite{2020cg, 2022hao, 2023hunt, 2024cavallo}). While the known clusters served to validate the performance of the HDBSCAN algorithm in this region,
their detailed multiwavelength analysis will be addressed in a future study.

\subsection{ESFERA 1}
\label{analysisEsf1}

Within the subregion defined by $320^{\circ} \leq l \leq 321^{\circ}$ and $-1^{\circ} \leq b \leq +1^{\circ}$, we analyzed a sample of 53076 stars. Using the HDBSCAN algorithm with \texttt{min\_samples} = \texttt{min\_cluster\_size} = 10, we identified an open cluster comprising 16 members. The mean astrometric parameters for this cluster are: $\overline{\alpha} = 227^{\circ}.323 \pm 0^{\circ}.028$, $\overline{\delta} = -58^{\circ}.427 \pm 0^{\circ}.011$, $\overline{\mu_{\alpha} \cos \delta} = -3.46 \pm 0.04$ mas yr$^{-1}$, $\overline{\mu_{\delta}} = -2.86 \pm 0.03$ mas yr$^{-1}$, $\overline{\varpi} = 0.39 \pm 0.01$ mas, and $\overline{d} = 2452 \pm 59$ pc.

\begin{figure*} 
\centering
\includegraphics[width=2\columnwidth]{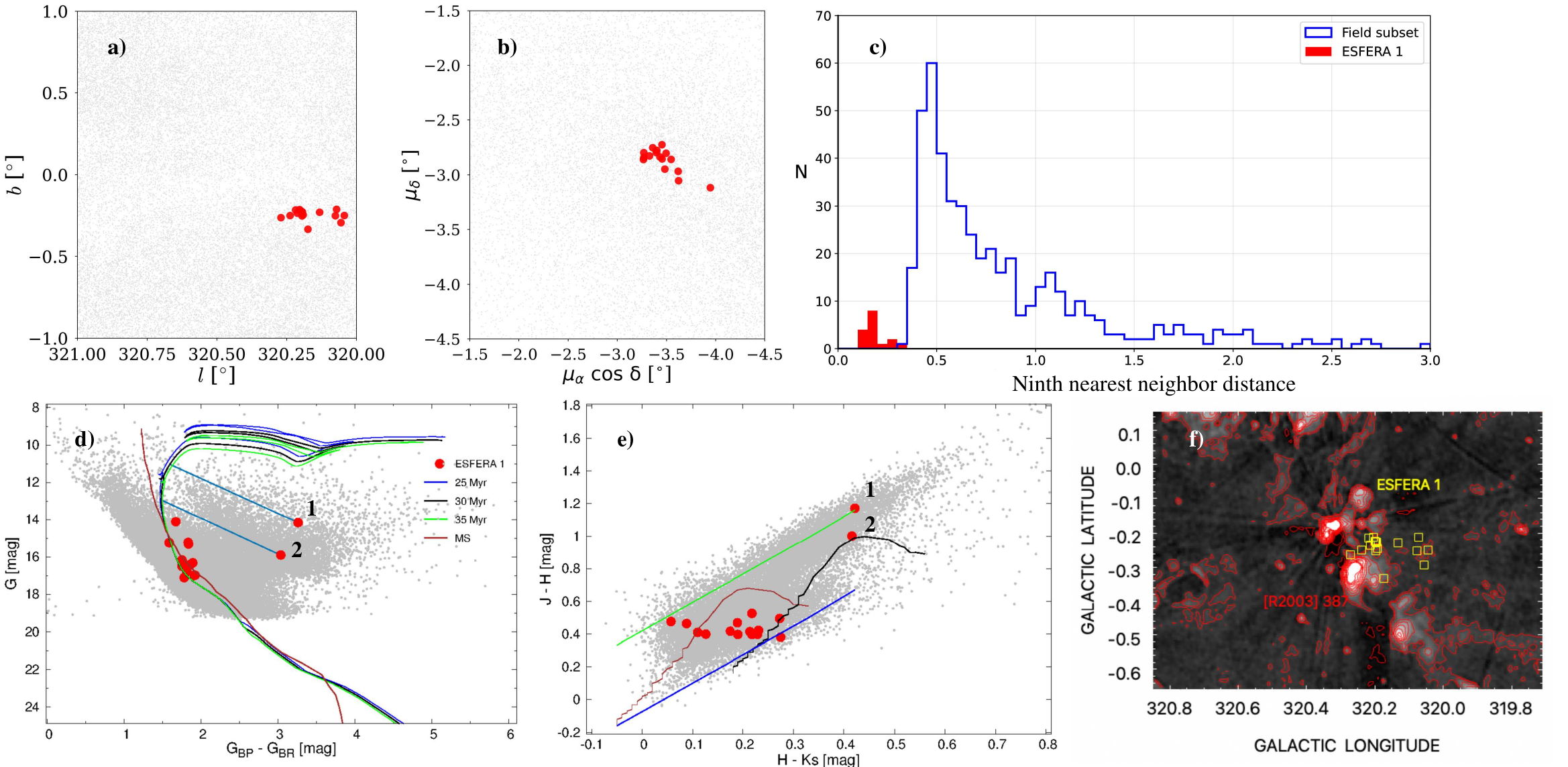}
\caption{Multi-parametric analysis of the open cluster \textit{ESFERA\,1}. (a) Spatial distribution in Galactic coordinates and (b) Vector Point Diagram (VPD) of the candidate members (red filled circles) compared to the field population (light gray dots). (c) Histogram of the 9th-nearest neighbor distances for \textit{ESFERA\,1} members (red) and a control subset of 500 local field stars (blue). (d) Optical Color-Magnitude Diagram (CMD); brown curve represents the Main Sequence (MS) fitted to a distance of 2452 pc and $A_V = 4$ mag. The black curve denotes the best-fit PARSEC v1.2S isochrone, while blue and green curves represent the $\sim$ 15$\%$ age uncertainties. (e) Near-infrared (NIR) CMD, where black and brown curves indicate the MS shifted according to the adopted distance, with and without interstellar reddening, respect. (f) Spatial projection of candidate members onto the SUMSS 843 MHz radio image. Red contours trace the HII emission of [R2003] 387 \citep{Murray10}; contour levels are 5 ($\sim 5\sigma_{\text{rms}}$), 10, 50, 100, 150, 200, 250, and 300 mJy beam$^{-1}$.}
\label{fig:Esf1FINAL}
    \end{figure*}

For \textit{ESFERA\,1}, we find $\sqrt{\sigma^{2}_{\mu_{\alpha} \cos \delta} + \sigma^{2}_{\mu_{\delta}}} = 0.41$ mas yr$^{-1}$ and $r_{50} = 2.08$ pc. Figure \ref{fig:Esf1FINAL}c shows that the members of $ESFERA\,1$ are more densely grouped than a subset of 500 local field stars, thus fulfilling the three conditions described in Section~\ref{candidates}.

The stars of this open cluster identified as \#1 and \#2 in Figs. \ref{fig:Esf1FINAL}d-e correspond to {\it Gaia} DR3 5880039205832933632 and {\it Gaia} DR3 5880038862235534720, respectively. Their $Q_{NIR}$ values of $0.45$ and $0.30$ suggest that both are likely early-type stars. By de reddening these stars using the $A_G/A_V = 0.789$ reddening law \citep{Wang19} they can be placed accurately on the 30\,Myr isochrone. Their derived mass and temperature values are consistent with spectral types B$^+$, possibly B3\,V and B4\,V, respectively. While these two potential members of \textit{ESFERA\,1} appear significantly reddened, their positions in the IR color-color diagram (see Fig. \ref{fig:Esf1FINAL}e) consistent with normally reddened stars. The other 14 potential members of the \textit{ESFERA\,1} open cluster, based on their placement along the isochrone, appear to be early-type stars ranging from B$^+$ to A$^+$. However, the Lyman-continuum photons emitted by these stars lack the necessary flux to ionize the interstellar medium and generate the HII region observed in this area. Furthermore, while the [R2003] 387 HII region \citep{Murray10} appears spatially coincident with \textit{ESFERA\,1}, its distance remains unknown. Another HII region, G\,320.252$-$00.332, was also investigated; however, its near-kinematic distance of 4.7 kpc \citep{Anderson14} confirms that it is not physically associated with the cluster. An average visual extinction of $A_v = 4.0$ mag was adopted for the cluster, as discussed in Sect. \ref{physical}. The complete list of members for \textit{ESFERA\,1}, including their astrometric parameters and multiwavelength photometry, is provided in Tables~\ref{table:astroesf1} and \ref{table:esfera1p}, respectively.

\subsection{ESFERA 2}
\label{analysisEsf2}

\begin{figure*}
  \centering
\includegraphics[width=2\columnwidth]{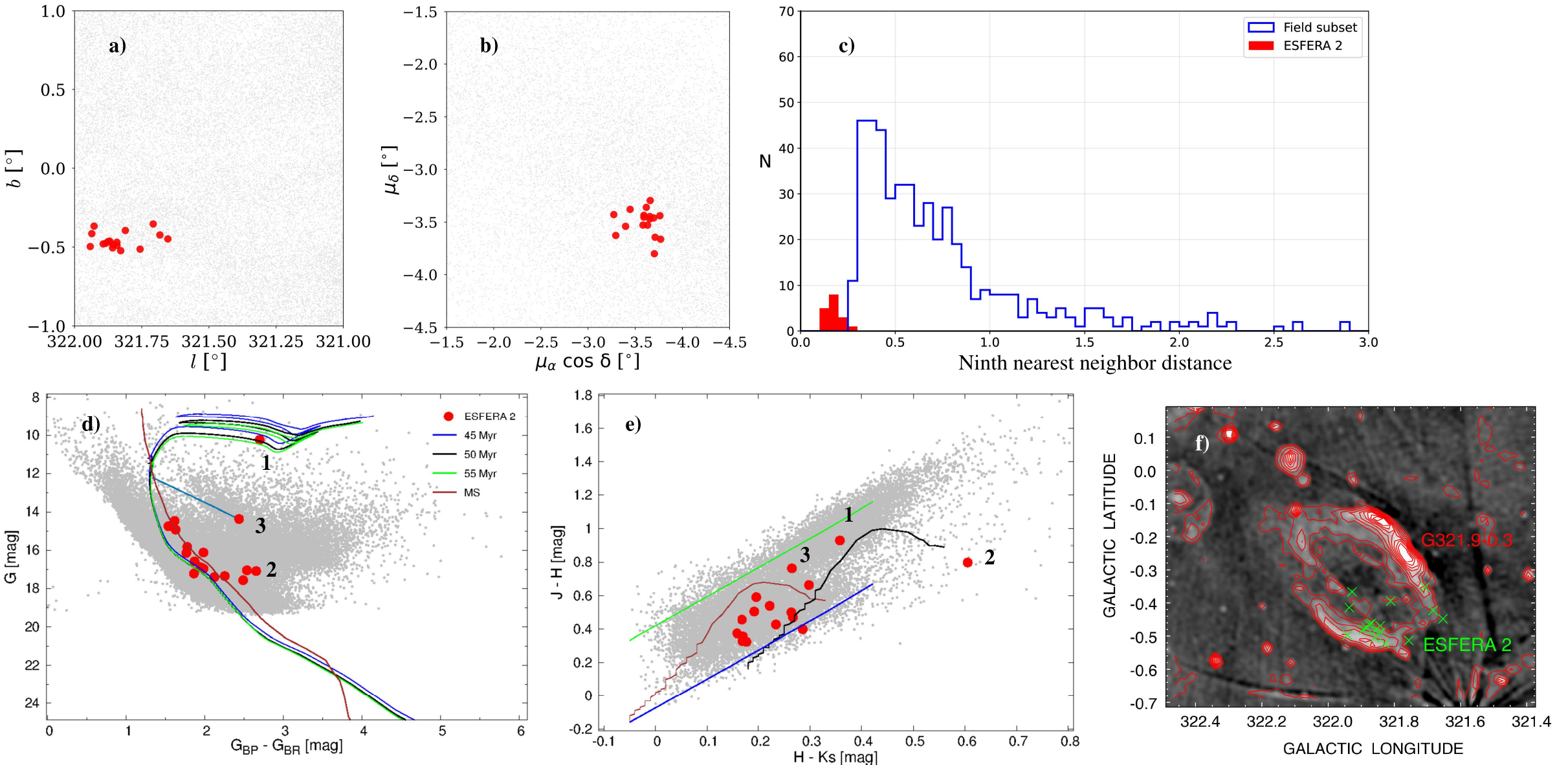}
\caption{Multi-parametric analysis of the open cluster \textit{ESFERA\,2}. Symbols as in Fig~\ref{fig:Esf1FINAL}. In Fig (d) the MS is fitted to a distance of 1907 pc and $A_V = 3.5$ mag; (f) Red contours trace the HII emission of G321.9$-$0.3 SNR \citep{Ball25}; contour levels are 2 ($\sim 3\sigma_{\text{rms}}$), 5, 10, 15, 20, 25, 30, 35 and 40 mJy beam$^{-1}$.}
\label{fig:Esf2FINAL}
 \end{figure*}

Within the subregion defined by $321^{\circ} \leq l \leq 322^{\circ}$ and $-1^{\circ} \leq b \leq +1^{\circ}$, we analyzed a sample of 51327 stars. Using \texttt{min\_samples} = \texttt{min\_cluster\_size} = 10, we identified an open cluster comprising 17 members. Its mean astrometric parameters are:
$\overline{\alpha} = 230^{\circ}.220 \pm 0^{\circ}.036$,
$\overline{\delta} = -57^{\circ}.737 \pm 0^{\circ}.013$,
$\overline{\mu_{\alpha} \cos \delta} = -3.59 \pm 0.04$ mas yr$^{-1}$, 
$\overline{\mu_{\delta}} = -3.50 \pm 0.03$ mas yr$^{-1}$, 
$\overline{\varpi} = 0.52 \pm 0.01$ mas, and 
$\overline{d} = 1907 \pm 32$ pc.
For \textit{ESFERA\,2}, we obtain a proper motion dispersion of $\sqrt{\sigma^{2}_{\mu_{\alpha} \cos \delta} + \sigma^{2}_{\mu_{\delta}}} = 0.51$ mas yr$^{-1}$ and a half-mass radius of $r_{50} = 2.16$ pc. As shown in Fig.~\ref{fig:Esf2FINAL}c, \textit{ESFERA\,2} is significantly denser than a representative subset of 500 local field stars, confirming its cluster nature.

The stars identified as \#1 and \#2 in Figs.~\ref{fig:Esf2FINAL}d-e correspond to {\it Gaia} DR3 5883123679555965184 and {\it Gaia} DR3 5883127935846172416, respectively. Star \#1 yields a $Q_{NIR}$ value of 0.32, which initially suggests an early-type main-sequence star. However, its position relative to the fitted isochrone in the CMD indicates that it has already evolved past the main-sequence turn-off point. Consequently, the appropriate luminosity class for this source is likely that of a giant (LC III) or supergiant (LC I). This classification is consistent with the estimated cluster age of 50\,Myr, derived from the evolutionary analysis of its probable members. Star \#2, as shown in Fig. \ref{fig:Esf2FINAL}e, is located to the right of the reddening vector for O-type dwarf stars and yields a $Q_{NIR}$ value of -0.23. This region of the $JHK$ diagram is typically populated by YSOs, such as Herbig Ae/Be and T Tauri stars, which exhibit near-infrared excess emission due to circumstellar disks \citep{Lada92}. To investigate this YSO candidacy, we cross-matched the source with the WISE catalogue to analyze its photometric flux according to the classification scheme described by \citet{Koenig14}. However, {\it Gaia} DR3 5883127935846172416 was not detected in the WISE survey, leaving its YSO status unconfirmed. Finally, star \#3 (Figs. \ref{fig:Esf2FINAL}d-e) corresponds to {\it Gaia} DR3 5883133128483745408. It presents a $Q_{NIR}$ value of 0.31, nearly identical to that of star \#1. In the infrared color-color diagram, it is similarly positioned near the giant reddening vector. In the {\it Gaia} CMD, star \#3 appears highly reddened, suggesting it is likely an evolved late-type B star.  The visual extinction for the cluster was estimated at $A_v = 3.5$ mag (see Sect.~\ref{physical}). The supernova remnant G\,321.9$-$0.3 \citep{Green19} is also located in the vicinity of \textit{ESFERA\,2} (see Fig.~\ref{fig:Esf2FINAL}f). However, multiple factors suggest this is a line-of-sight coincidence. 
A detailed discussion on this topic is provided in Sect.~\ref{discussion}. The complete list of members for \textit{ESFERA\,2}, including their astrometric parameters and multiwavelength photometry, is provided in Tables~\ref{table:astroesf2} and \ref{table:esfera2p}, respectively.

\subsection{ESFERA 3}
\label{analysisEsf3}

Within the subregion $322^{\circ} \leq l \leq 323^{\circ}$ and $-1^{\circ} \leq b \leq +1^{\circ}$, we analyzed a sample of 45535 stars. Setting \texttt{min\_samples} = \texttt{min\_cluster\_size} = 10, we identified an open cluster with 20 members. Its mean astrometric parameters are:
$\overline{\alpha} = 230^{\circ}.079 \pm 0^{\circ}.032$,
$\overline{\delta} = -57^{\circ}.189 \pm 0^{\circ}.020$,
$\overline{\mu_{\alpha} \cos \delta} = -2.22 \pm 0.05$ mas yr$^{-1}$, 
$\overline{\mu_{\delta}} = -2.15 \pm 0.04$ mas yr$^{-1}$, 
$\overline{\varpi} = 0.50 \pm 0.01$ mas, and 
$\overline{d} = 1939 \pm 34$ pc.

\begin{figure*} 
   \centering
\includegraphics[width=2\columnwidth]{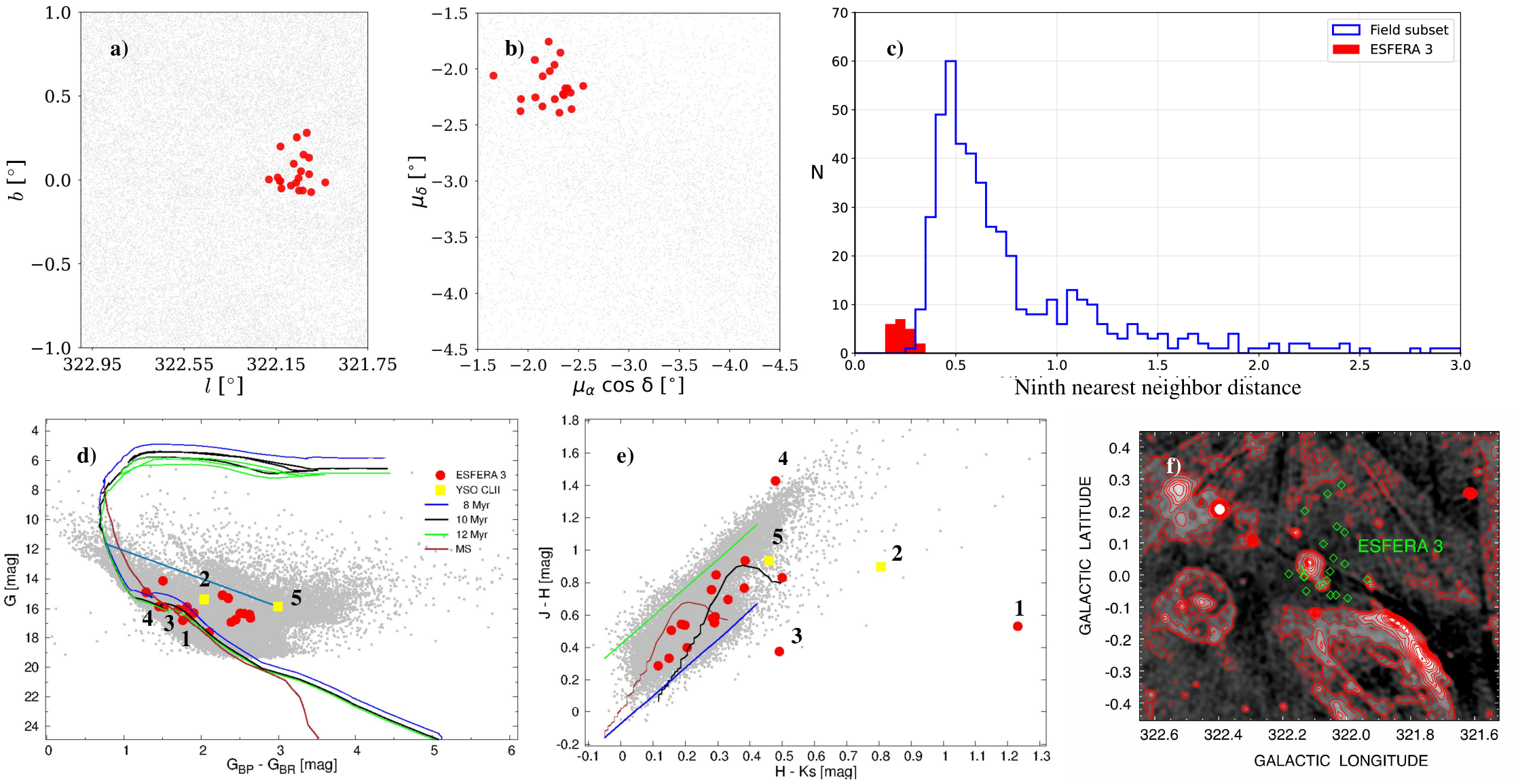}
\caption{Multi-parametric analysis of the open cluster \textit{ESFERA\,3}. Symbols as in Fig~\ref{fig:Esf1FINAL} and yellow square for Class II YSO. In Fig (d) the MS is fitted to a distance of 1939 pc and $A_V = 2.5$ mag; (f) Red contours trace the same levels of Fig.~\ref{fig:Esf2FINAL} f).}
\label{fig:Esf3FINAL}
    \end{figure*}

\begin{figure}[!t] 
  \centering
\includegraphics[width=\columnwidth]{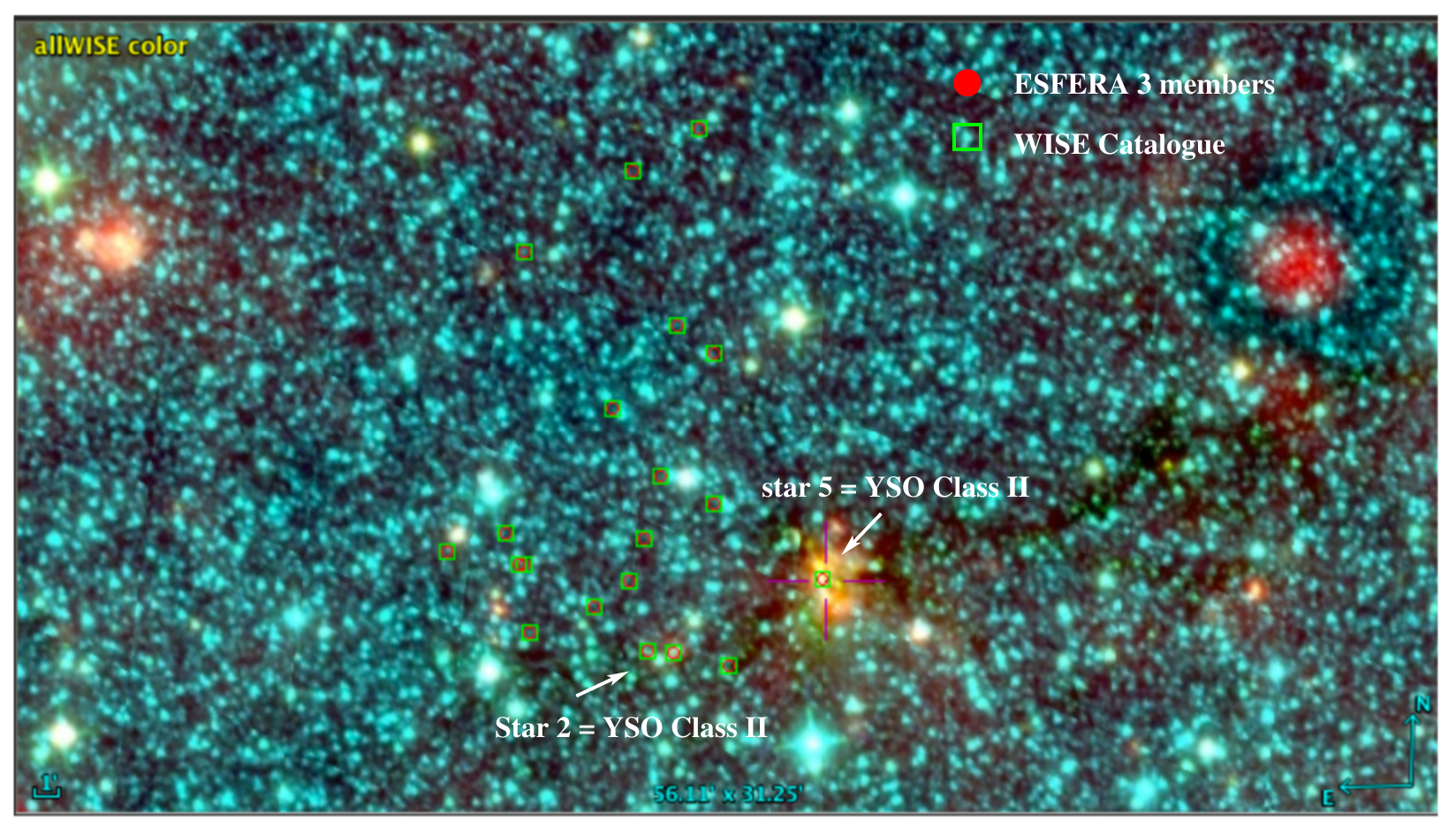}
\caption{AllWISE color composite image of the \textit{ESFERA\,3} region (l = 320$^{\circ}$.0, b = 0$^{\circ}$.0). Red circles identify the cluster members, while green squares represent cataloged YSOs. Two probable Class II YSO members of the \textit{ESFERA\,3} cluster are indicated.}
\label{fig:Esf3YSO}
   \end{figure}

For \textit{ESFERA\,3}, we obtain a proper motion dispersion of $\sqrt{\sigma^{2}_{\mu_{\alpha} \cos \delta} + \sigma^{2}_{\mu_{\delta}}} = 0.52$ mas yr$^{-1}$ and a half-mass radius of $r_{50} = 3.57$ pc. As illustrated in Fig.~\ref{fig:Esf3FINAL}c, \textit{ESFERA\,3} is more densely populated than a representative subset of 500 local field stars, supporting its classification as a physical cluster.
\textit{ESFERA\,3} is the youngest open cluster in our sample, with an estimated age of only 10\,Myr. In Fig.~\ref{fig:Esf3FINAL}e, the presence of reddened stars is clearly evident, specifically stars \#1 ({\it Gaia} DR3 5883216622652247808), \#2 ({\it Gaia} DR3 5883215003418857472), and \#3 ({\it Gaia} DR3 5883219440150823424). Stars \#1 and \#3 are recorded in the 2MASS catalog with a quality flag ($Q_{fl}$) of `AAU'. This indicates that while the $J$ and $H$ band detections are of high photometric quality, the $K_s$ band measurement is an upper limit, likely due to undefined photometric errors in that filter. In the CMD (Fig.~\ref{fig:Esf3FINAL}d), these stars are positioned close to each other on the fitted isochrone, allowing us to derive spectral types of F$^-$ for star \#1 and A$^+$ for star \#3. Star \#2 presents a $Q_{NIR}$ value of $-0.47$, and its position in the IR color-color diagram exhibits a significant near-infrared excess. Our investigation of the WISE survey, utilizing the classification scheme by \citet{Koenig14}, identifies this source as a Class II YSO. These objects are typically characterized by the presence of a protoplanetary disk where circumstellar dust reprocesses stellar radiation, creating an infrared excess, even as the primordial envelope begins to dissipate \citep{Allen04}. Regarding star \#4 ({\it Gaia} DR3 5883232359411945472), its position in the CMD (see Fig.~\ref{fig:Esf3FINAL}d) relative to the fitted isochrone corresponds to temperature and mass values consistent with an A$^+$ spectral type. However, its placement in the IR color-color diagram (see Fig.~\ref{fig:Esf3FINAL}e) is likely an artifact of poor photometric quality, as indicated by its $Q_{fl}$ flag of `AUU'. This flag denotes that both the $H$ and $K_s$ magnitudes are upper limits, rendering its infrared colors unreliable. Star \#5 ({\it Gaia} DR3 5883194941656356480) is the most heavily reddened star in Fig.~\ref{fig:Esf3FINAL}d. It presents a $Q_{NIR}$ value of 0.15, characteristic of an early main-sequence star, and its position in the Fig.~\ref{fig:Esf3FINAL}e is consistent with a late spectral type (with a high-quality $Q_{fl}$ of `AAA'). This source have been identified as a Class II YSO, cataloged as AGAL G321.934$-$00.006 in the ATLASGAL survey of massive cold dust clumps \citep{Wienen15}. It has a reported heliocentric distance of 2.14 kpc, which is in close agreement with the estimated distance for \textit{ESFERA\,3}. We independently verified this classification through an analysis of the WISE catalogue and both YSOs are indicated in the allWISE color image Fig~\ref{fig:Esf3YSO}. Finally, the stars spatially coincident with the HI emission --a possible HII region-- possess spectral types A$^-$, F$^-$, and B$^+$. Consequently, none of these stars provide sufficient ionizing flux to ionize the surrounding interstellar medium. The complete list of members for \textit{ESFERA\,3}, including their astrometric parameters and multiwavelength photometry, is provided in Tables~\ref{table:astroesf3} and \ref{table:esfera3p}, respectively.

\subsection{ESFERA 4}
\label{analysisEsf4}

\begin{figure*} 
   \centering
\includegraphics[width=2\columnwidth]{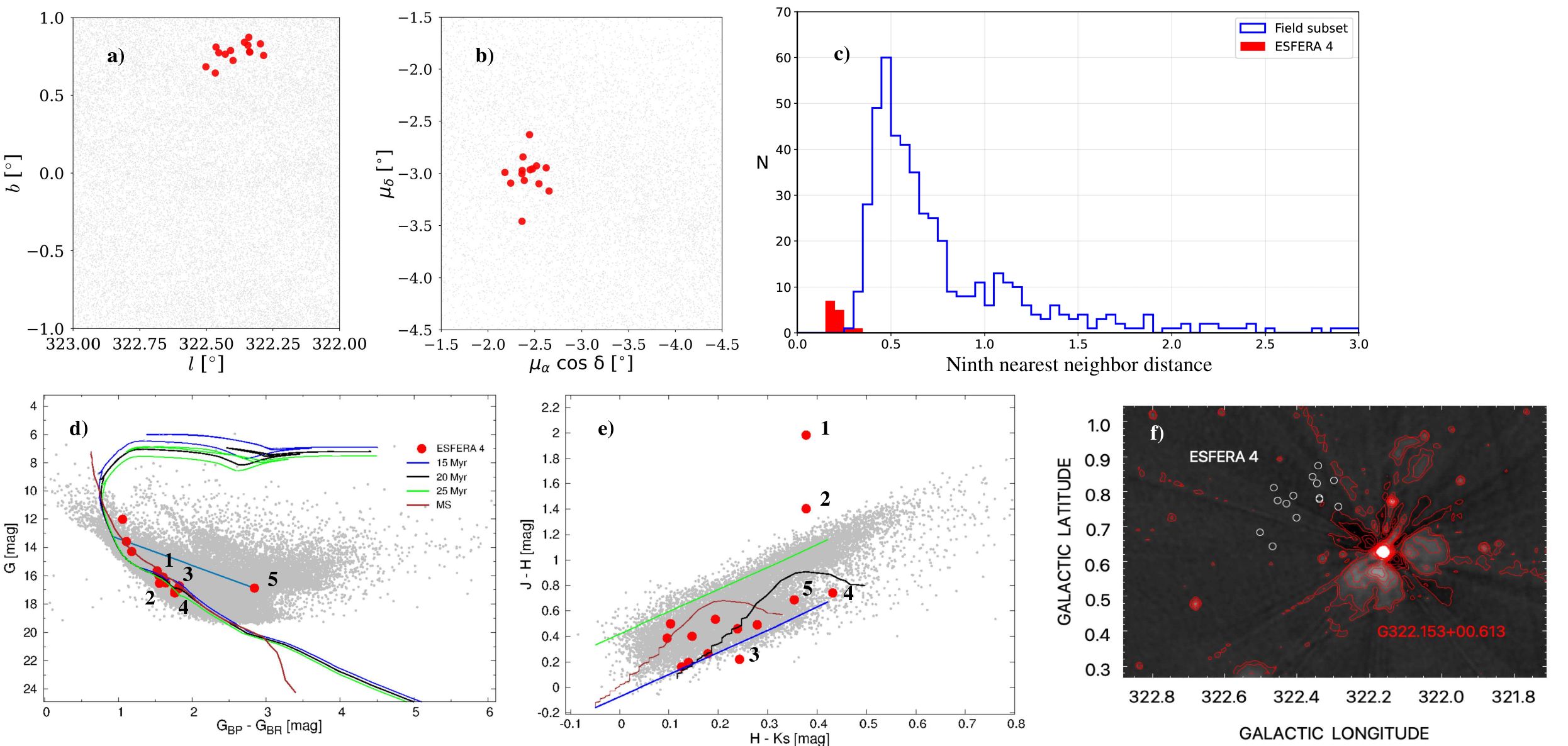}
\caption{Multi-parametric analysis of the open cluster \textit{ESFERA\,4}. Symbols as in Fig.~\ref{fig:Esf1FINAL}. In Fig (d) the MS is fitted to a distance of 1648 pc and $A_V = 2.5$ mag; (f) Red contours trace the H\,II emission of G\,322.153$+$00.613 \citep{Anderson14}, contour levels are: -60, -40, -20 and -10 mJy beam$^{-1}$ for the negative emission and 5 ($\sim 3\sigma_{\text{rms}}$), 10, 30, 50, 70, 100, 150, 200, 250 and 300 mJy beam$^{-1}$ for the positive emission.}
\label{fig:Esf4FINAL}
    \end{figure*}

In addition to \textit{ESFERA\,3}, within the same subregion ($322^{\circ} \leq l \leq 323^{\circ}$, $-1^{\circ} \leq b \leq +1^{\circ}$) and using \texttt{min\_samples} = \texttt{min\_cluster\_size} = 10, we identified another open cluster consisting of 14 members. Its mean astrometric parameters are:
$\overline{\alpha} = 229^{\circ}.860 \pm 0^{\circ}.038$,
$\overline{\delta} = -56^{\circ}.399 \pm 0^{\circ}.011$,
$\overline{\mu_{\alpha} \cos \delta} = -2.43 \pm 0.03$ mas yr$^{-1}$, 
$\overline{\mu_{\delta}} = -3.01 \pm 0.05$ mas yr$^{-1}$, 
$\overline{\varpi} = 0.59 \pm 0.01$ mas, and 
$\overline{d} = 1648 \pm 24$ pc.
For \textit{ESFERA\,4}, we find a proper motion dispersion of $\sqrt{\sigma^{2}_{\mu_{\alpha} \cos \delta} + \sigma^{2}_{\mu_{\delta}}} = 0.61$ mas yr$^{-1}$ and a half-mass radius of $r_{50} = 1.90$ pc. As shown in Fig.~\ref{fig:Esf4FINAL}c, \textit{ESFERA\,4} is denser than a representative subset of 500 local field stars.

Stars \#1 ({\it Gaia} DR3 58863349440504533248) and \#2 ({\it Gaia} DR3 58863363386248420736) possess 2MASS photometric $Q_{fl}$ of `AUU', indicating that their resulting infrared colors are unreliable, similar to star \#4 in \textit{ESFERA\,3}. The calculated $Q_{NIR}$ values of 1.34 and 0.76 for sources number \#1 and \#2 are likely non-physical artifacts. When plotted on the 20\,Myr isochrone, the positions of these two stars suggest they are late-type objects, with estimated spectral types of F$^-$ and F$^+$, respectively. Stars \#3 ({\it Gaia} DR3 58863257425123101568) and \#4 ({\it Gaia} DR3 58863352429801298816) are the oldest members of \textit{ESFERA\,4}, both characterized by a G$^-$ spectral type. Regarding their 2MASS $JHK_s$ fluxes, the former has a $Q_{fl}$ of `AAU' and a $Q_{NIR}$ of $-0.19$, while the latter has a $Q_{fl}$ of `AAA' and a $Q_{NIR}$ of $0.00$. The positions of both stars in the IR color-color diagram are typical for their respective spectral types. Star \#5 ({\it Gaia} DR3 58863352567240269312) is the most heavily reddened member of the cluster. By fitting the isochrone with the $A_G/A_V = 0.789$ reddening law, we derive a most probable spectral type of B$^+$. Interestingly, its placement in the IR color-color diagram is somewhat anomalous, as it lies within the region typically occupied by late-type main-sequence stars. However, given that the photometric quality in all three bands is excellent (`AAA') and the $Q_{NIR}$ value is $0.08$ --consistent with a standard stellar atmosphere-- this displacement is likely a result of the high extinction affecting the source rather than an intrinsic near-infrared excess. Regarding the interstellar medium (see Fig.~\ref{fig:Esf4FINAL}f), this open cluster is located in proximity to the HII region G322.153$+$00.613, although their spatial positions do not overlap. The complete list of members for \textit{ESFERA\,4}, including their astrometric parameters and multiwavelength photometry, is provided in Tables~\ref{table:astroesf4} and \ref{table:esfera4p}, respectively.

\subsection{ESFERA 5}
\label{analysisEsf5}

\begin{figure*}
   \centering
\includegraphics[width=2\columnwidth]{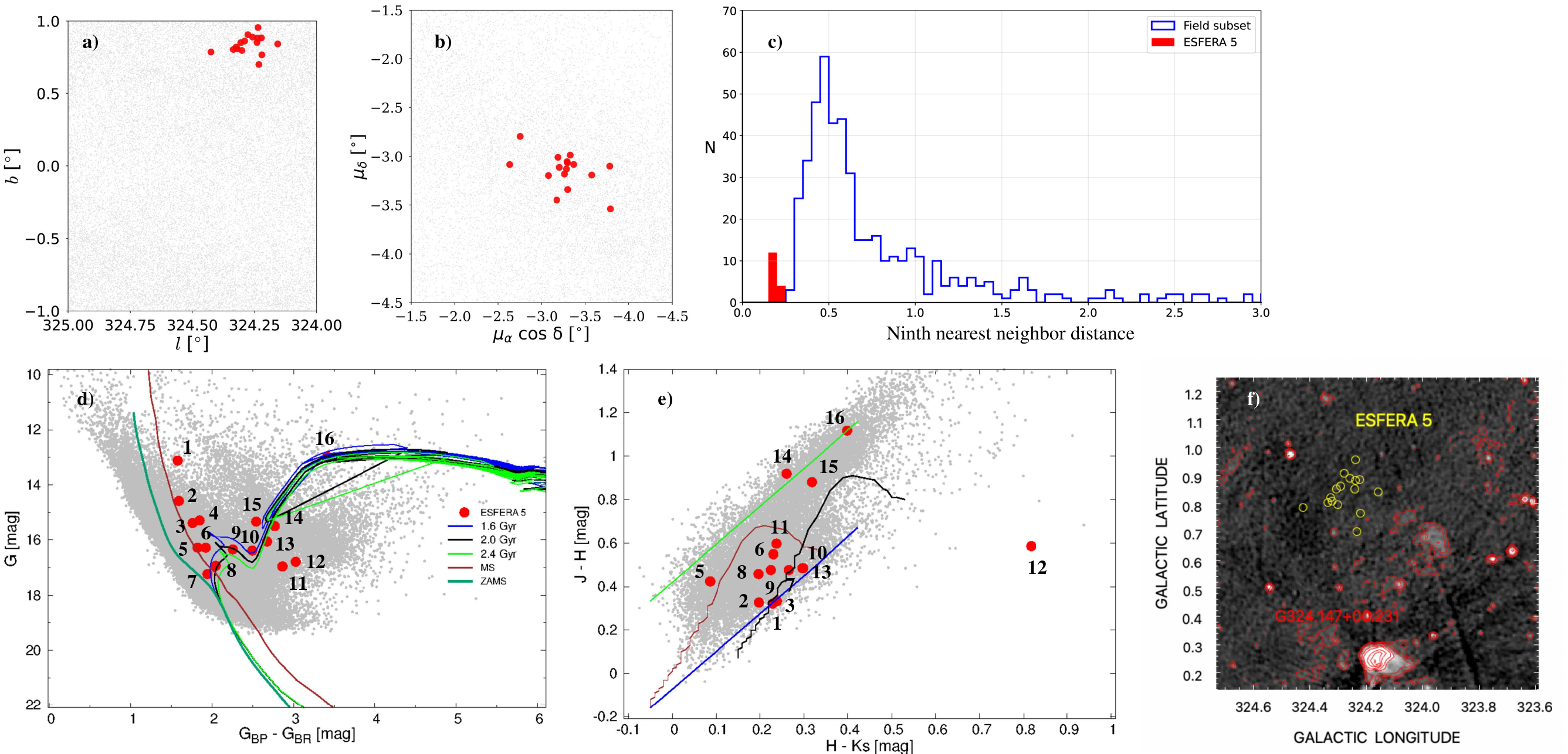}
\caption{Multi-parametric analysis of the open cluster \textit{ESFERA\,5}. Symbols as in Fig~\ref{fig:Esf1FINAL}. In Fig (d) the MS is fitted to a distance of 2417 pc and $A_V = 3.4$ mag; (f) Red contours indicate the H\,II emission of G324.147$+$00.231 \citep{Anderson14}. Contour levels are 2 ($\sim 5 \sigma_{\text{rms}}$), 5, 10, 30, 50, 100, 150, 200, 250 and 300 mJy beam$^{-1}$.}
\label{fig:Esf5FINAL}
    \end{figure*}

Within the subregion $324^{\circ} \leq l \leq 325^{\circ}$ and $-1^{\circ} \leq b \leq +1^{\circ}$, we analyzed a sample of 52417 stars. By increasing the clustering threshold to \texttt{min\_samples} = \texttt{min\_cluster\_size} = 15, we identified an open cluster consisting of 16 members. Its mean astrometric parameters are:
$\overline{\alpha} = 232^{\circ}.601 \pm 0^{\circ}.029$,
$\overline{\delta} = -55^{\circ}.721 \pm 0^{\circ}.013$,
$\overline{\mu_{\alpha} \cos \delta} = -3.27 \pm 0.07$ mas yr$^{-1}$, 
$\overline{\mu_{\delta}} = -3.15 \pm 0.04$ mas yr$^{-1}$, 
$\overline{\varpi} = 0.40 \pm 0.01$ mas, and 
$\overline{d} = 2417 \pm 49$ pc.
For \textit{ESFERA\,5}, we obtain a proper motion dispersion of $\sqrt{\sigma^{2}_{\mu_{\alpha} \cos \delta} + \sigma^{2}_{\mu_{\delta}}} = 0.41$ mas yr$^{-1}$ and a half-mass radius of $r_{50} = 2.41$ pc. As illustrated in Fig.~\ref{fig:Esf5FINAL}c,  \textit{ESFERA\,5} is more densely populated than a representative subset of 500 local field stars

This is a particularly noteworthy open cluster. At 2\,Gyr, it is significantly older than the other clusters identified in our study. Furthermore, it is the only one among the {\bf five} discovered clusters that contains potential Blue Straggler Stars (BSS) among its probable members. These candidates, labeled \#1 ({\it Gaia} DR3 5883844512499441280) through \#6 ({\it Gaia} DR3 5883838911861950720), are located above and blueward of the Main Sequence turn-off point. They represent a population of rejuvenated stars, whose presence points toward an interesting dynamical history that will be further analyzed in the Sect. \ref{discussion}. Stars \#7 ({\it Gaia} DR3 5883837915429150592) and \#8 ({\it Gaia} DR3 5883841832439453056) are the only sources fitted on the main sequence with a luminosity class V (see Fig.~\ref{fig:Esf5FINAL}d). The remaining sources, numbered \#9 ({\it Gaia} DR3 5883845165334464256) through \#16 ({\it Gaia} DR3 5883837086485049088) are potential giants; in Fig.~\ref{fig:Esf5FINAL}e, stars \#9, \#10 ({\it Gaia} DR3 5883844100182546944), \#11 ({\it Gaia} DR3 5883840698568446208), and \#13 ({\it Gaia} DR3 5883844375059980800) occupy the region typical of late-type stars, while stars \#14 ({\it Gaia} DR3 5883843992792751232), \#15 ({\it Gaia} DR3 5883844993535795840), and \#16 are consistent with the giant branch. Notably, star \#4 ({\it Gaia} DR3 5883842210396958720) has no reported magnitudes in the near-infrared $J,H,K$ bands and star \#12 ({\it Gaia} DR3 5883840732928203776) has a $Q_{fl}$ = `AUU'. The last result, likely explains its significant displacement from the sequence where stars with standard interstellar extinction are expected to lie. The complete list of members for \textit{ESFERA\,5}, including their astrometric parameters and multiwavelength photometry, is provided in Tables~\ref{table:astroesf5} and \ref{table:esfera5p}, respectively.

\begin{table*}
\caption{Open clusters $ESFERA$ discovered in this study for the galactic region $320^{\circ} \leq l \leq 325^{\circ}$,  $-1^{\circ} \leq b  \leq +1^{\circ}$}
\label{table:generalparam}
\centering
\fontsize{8pt}{9pt}\selectfont
\setlength{\tabcolsep}{3pt}
\begin{tabular}{c c c c c c c c c c} 
    \toprule
$ID$ & $GLON$ & $GLAT$ & $\alpha$ &	$\delta$ &	$\mu_{\alpha}\rm{cos}\delta$ &	$\mu_{\delta}$ & Dist$_{pl}$ &	$A_v$ &	$AGE$ \\
    \midrule 
     & $[\circ]$ & $[\circ]$ & $[\circ]$ & $[\circ]$ & [mas yr$^{-1}$] & [mas yr$^{-1}$] & $[pc]$ & $[mag]$ & $[Myr]$  \\ 
    \midrule 
    \begin{filecontents*}{ParamESFERAS1.csv}
ID,GLON,GLAT,AR,DEC,pmAR,pmDEC,Dpl,Av,Age
1,320.167$\pm$0.017,-0.246$\pm$0.008,227.323$\pm$0.028,-58.427$\pm$0.011,-3.46$\pm$0.04,-2.86$\pm$0.03,2452$\pm$59,4.0$\pm$0.4,30$\pm$5
2,321.833$\pm$0.020,-0.456$\pm$0.012,230.220$\pm$0.036,-57.737$\pm$0.013,-3.59$\pm$0.04,-3.50$\pm$0.03,1907$\pm$32,3.5$\pm$0.3,50$\pm$5
3,322.065$\pm$0.013,0.045$\pm$0.023,230.079$\pm$0.032,-57.189$\pm$0.020,-2.22$\pm$0.05,-2.15$\pm$0.04,1939$\pm$34,2.5$\pm$0.2,10$\pm$2
4,322.387$\pm$0.018,0.776$\pm$0.016,229.860$\pm$0.038,-56.399$\pm$0.011,-2.43$\pm$0.03,-3.01$\pm$0.05,1648$\pm$24,2.5$\pm$0.2,20$\pm$5
5,324.274$\pm$0.015,0.838$\pm$0.015,232.601$\pm$0.029,-55.306$\pm$0.013,-3.27$\pm$0.07,-3.15$\pm$0.04,2417$\pm$49,3.4$\pm$0.3,2000$\pm$400
\end{filecontents*}
\csvreader[
        head to column names, 
        separator=comma,
        late after line=\\
    ]{ParamESFERAS1.csv}{}{
        \ID & \GLON & \GLAT & \AR & \DEC & \pmAR & \pmDEC & \Dpl & \Av & \Age  
    }
    \bottomrule 
\end{tabular}
\end{table*}    

\section{Discussion}
\label{discussion}
Our analysis reveals some diverse population of open clusters tracing the Carina-Sagittarius arm within $320^{\circ} \leq l \leq 325^{\circ}$. We find a diversity in morphology, with structures varying from compact, such as \textit{ESFERA\,4} ($r_{50} < 2$\,pc), to more extended, such as \textit{ESFERA\,3} ($r_{50} > 3.5$\,pc). Furthermore, the clustering algorithm effectively separated structures projected along the same line of sight but located at different distances, such as \textit{ESFERA\,3} ($\sim 1.9$\,kpc) and \textit{ESFERA\,4} ($\sim 1.6$\,kpc), providing a clearer view of the stellar distribution at different depths of the spiral arm. Figure~\ref{fig:HistogOCs} reveals the density drops by more than 50\% compared to adjacent regions. The discovery of the five ESFERA clusters in this specific 'gap' significantly improves the census of the stellar population in this under-sampled section of the Galactic disc. The robustness of our methodology stems from the HDBSCAN algorithm's ability to operate in a high-dimensional parameter space, which allows for the effective segregation of stellar populations that appear blended in traditional proper motion or spatial projections. By simultaneously processing five-dimensional astrometric data ($\alpha, \delta, \mu_{\alpha}\cos\delta, \mu_{\delta}, \varpi$), our approach successfully identifies distinct stellar groups that share nearly identical proper motions but are clearly separated by their parallax values. This high-dimensional clustering capability is critical for minimizing field star contamination and enables the detection of sparse clusters that would remain hidden in traditional 2D or 3D searches, particularly within the high-density and high-extinction environments of the Galactic plane.

\subsection{ESFERA 1}
The estimated age of 30\,Myr for \textit{ESFERA\,1}, derived from the isochrone fitting, strongly supports the conclusion that the cluster is not physically associated with the HII region [R2003] 387. At an age of 30\,Myr, any massive O-type stars which are essential for providing the ionizing Lyman-continuum flux required to maintain an HII region would have already evolved off the main sequence or ended their life cycles as supernovae. The most massive members identified in this study, classified as B$^+$ (probably B3 V and B4 V spectral types), are perfectly consistent with a 30\,Myr population; however, their ionizing photon rate ($\log N_{LyC}$) is several orders of magnitude below the threshold needed to sustain the observed radio emission in the region. Furthermore, while stars \#1 and \#2 exhibit significant reddening, their alignment with the standard reddening vector in the $JHK_s$ IR color-color diagram suggests that this extinction is primarily interstellar rather than circumstellar. This indicates that \textit{ESFERA\,1} has already cleared its natal gas and dust, a process typically completed within the first 10\,Myr of a cluster's life. Consequently, the spatial coincidence between \textit{ESFERA\,1} and the [R2003] 387 HII region is likely a chance alignment along the line of sight. This is analogous to our findings for G320.252-00.332, where the near-kinematic distance of 4.7\,kpc \citep{Anderson14} clearly places it at a different Galactic location, further confirming that \textit{ESFERA\,1} is an independent stellar population.

\subsection{ESFERA 2, 3 and 4}
The spatial proximity of these new candidates to interstellar features provides further insights into their evolutionary history. For \textit{ESFERA\,2} the identification of Star \#2 as a potential YSO candidate based on its $JHK_s$ color excess presents a chronological discrepancy. While this excess suggests the presence of a circumstellar disk, such structures typically dissipate within 5–10 Myr, a timescale significantly shorter than the estimated 50 Myr age of the cluster. Since the source lacks a WISE counterpart to confirm a mid-infrared excess, its position in the IR color-color diagram might instead be explained by extreme localized extinction or, alternatively, it may represent a background YSO unrelated to the cluster. Another possibility is for star \#2 to be a classical Be star, where the infrared excess originates from a gaseous decretion disk rather than a primordial pre-main-sequence disk \citep{Zorec97}, which would be more consistent with the evolved nature of the cluster. Another relevant aspect regarding this cluster is its possible relationship with the SNR G321.9$-$0.3. Previous estimates place this remnant at a distance of approximately 5.5 kpc \citep{Case98}, which contrasts with our derived distance for the cluster. More importantly, the SNR is associated with the X-ray binary Circinus X-1, a system originating from a very massive progenitor. Given that \textit{ESFERA\,2} is 50 Myr old, any such massive stars would have exploded tens of millions of years ago. The presence of a visible SNR today --which typically lasts only $\sim10^5$ years-- clearly indicates a much more recent supernova event unrelated to the cluster's evolution.

A similar environmental analysis was conducted for \textit{ESFERA\,3} focusing on its association with local gas. It is noteworthy that while stars \#1, \#3, and \#4 (spectral types A and F) of this OC are located in regions with detectable HI emission, star \#2 is not. This contrast reinforces the classification of star \#2 as a more "primordial" or less evolved object compared to the other members. The A and F-type stars, despite their young age, appear to have already emerged from the densest molecular cores, whereas star \#2 remains tied to a region that has not yet transitioned to a diffuse atomic phase.

Regarding \textit{ESFERA\,4}, as noted in Sect.~\ref{analysisEsf4}, this cluster is located in proximity to the HII region G322.153$+$00.613, although their current spatial positions do not overlap. Furthermore, an analysis of the members' proper motions suggests that their locations were also not coincident at the time of cluster formation; since their motion is oriented toward the HII region (see Fig.~\ref{fig:mulmub}), it is unlikely that they share a common origin or that triggered star formation occurred between them.

\subsection{ESFERA 5}
 
The open cluster \textit{ESFERA\,5} presents an estimated age of $\sim 2$ Gyr, which is highly atypical for an open cluster situated near the Galactic plane. Open clusters in this region are generally expected to suffer rapid dynamical disruption \citet{Janes94} due to frequent encounters with Giant Molecular Clouds (GMCs) \citet{Spitzer58} and strong Galactic tidal forces \citet{Wielen85}. The longevity of this specific cluster, however, suggests a distinct survival mechanism. Its persistence up to 2 Gyr, despite being close to the disk, is further supported by the presence of BSS among its members. The stellar membership of \textit{ESFERA\,5} was robustly established via astrometric analysis of {\it Gaia} DR3 data. Following this, the BSS population was identified from the CMD by their position significantly above and blueward of the Main Sequence turn-off point, consistent with a rejuvenation of stars more massive than the cluster turn-off mass. The astrometric solution for these BSS candidates indicates a low RUWE value wich could initially suggest a lack of close binary activity, we must account for the fact that the sensitivity of RUWE to binarity decreases significantly at distances of $\sim$2.5 kpc \citep{2022Penoyre, 2024cas}. Consequently, we consider two potential formation channels for the BSS population in \textit{ESFERA\,5}. First, if these stars are indeed single objects as the astrometric solution suggests, their formation would likely be driven by direct stellar collisions in a high-density core environment. This would imply that \textit{ESFERA\,5} was significantly more massive and compact in its early stages. Alternatively, given the distance-related limitations of RUWE, we cannot rule out the mass transfer channel in primordial binary systems \citep{Rain24}. The presence of such a significant BSS population—regardless of the specific formation mechanism—remains a compelling indicator of the cluster's complex dynamical evolution over its 2 Gyr lifetime. Furthermore, the 2\,Gyr age of \textit{ESFERA\,5} denotes an evolved stellar population that lacks the capacity to ionize or mechanically disrupt the local interstellar medium. The HII region G324.147$+$00.231 \citep{Anderson14} indicated in Fig.~\ref{fig:Esf5FINAL}f is located at an angular distance of $\sim 38'$, which is approximately four times the cluster's estimated radius. This significant spatial offset, combined with the fact that its population cannot generate such HII region, confirms that G324.147$+$00.231 is a non$-$associated object, likely projected along a similar line of sight. Old open clusters are typically not considered primary tracers of recent star formation within spiral features, nevertheless, we included \textit{ESFERA\,5} in our analysis due to its unique characteristics and its absence in previous catalogs.

 The detection of \textit{ESFERA\,5} ($\log t \approx 9.3, A_V > 3, b = 0^{\circ}$.8) is particularly consistent with the selection functions described by 
\citet[see their Fig.~3]{Hunt2025}. According to their analysis, clusters in this age and extinction regime within the Galactic plane have a recovery probability below 15$\%$, highlighting the effectiveness of our high-purity regional search in identifying older, sparse populations that often elude global automated algorithms. Despite its estimated age of 2~Gyr, this cluster is particularly noteworthy for hosting a population of Blue Straggler Stars among its members. The identification of such a mature system, containing clear evidence of stellar interactions, provides valuable insights into the dynamical history and the long-term stellar population of the Carina–Sagittarius region.

The findings in the current $320^{\circ} \le l \le 325^{\circ}$ sector are consistent with, yet distinct from, previous investigations in the neighboring Carina region. Specifically, at $l = 316^{\circ}.8$, a multi-wavelength approach revealed a population of three embedded clusters associated with early-type stars and different classes of YSOs \citep{Corti23}. The presence of these objects at $l = 316^{\circ}.8$ provides a direct point of comparison for the \textit{ESFERA} sample, marking the transition from the active star-forming complexes of the Carina arm toward a more quiescent sector. The region spanning $l \in [320^{\circ}, 325^{\circ}]$ exhibits a significant deficit of young star-forming tracers, compared to the $l \approx 316^{\circ}$ area. We propose that this sector represents a clear observational example of a star-formation `valley' (or gap) resulting from the periodic fragmentation of the interstellar medium, as modeled by the magneto-gravitational instability \citet{1982el,2002kim}. 

In the context of the `beads on a string' morphology described by \cite{2010e}, our study area appears to function as the `string' or the low-density node connecting two major star-forming complexes (the `beads'): the dense Carina region at $l < 310^{\circ}$ and the Centaurus complex at $l > 330^{\circ}$. This spatial distribution suggests that large-scale gravitational and magnetic forces have effectively channeled gas away from this segment to feed the adjacent massive complexes. Consequently, the local surface density in this region has fallen below the critical threshold required to trigger collapse, maintaining it in a quiescent state compared to its immediate galactic surroundings. In this context, the discovery of the \textit{ESFERA} sample provides a crucial contribution to this `beads on a string' scenario. While the region $320^{\circ} \leq l \leq 325^{\circ}$ lacks the massive star-forming complexes found in Carina or Centaurus, our identification of five new open clusters --particularly the youngest systems like \textit{ESFERA\,3} with its associated YSOs-- demonstrates that star formation is not entirely suppressed in this `string' segment. Instead, it occurs in smaller, more isolated pockets where the local gas density managed to exceed the collapse threshold. This suggests that the `valley' is not a void, but rather a region of lower-efficiency star formation that requires high-precision surveys like ours to be fully characterized.

\vspace{-0.2cm}
\section{Conclusions}
This work bridges a significant observational gap in the Carina–Sagittarius Arm, specifically within the region $320^{\circ} \leq l \leq 325^{\circ}$, by conducting a systematic search for previously uncatalogued open clusters. To achieve this, we performed a comprehensive multiwavelength analysis, integrating both astrometric and photometric data from {\it Gaia} DR3 with Near- and Mid-Infrared photometry from 2MASS and WISE, and radio continuum data from SUMSS (843 MHz). This approach led to the discovery of five open clusters, all located within the Galactic plane ($-1^{\circ} \leq b \leq 1^{\circ}$). Our sample is primarily composed of four young systems with ages ranging between 10 and 50 Myr. Additionally, we identified \textit{ESFERA\,5} as a significantly older system, with an estimated age of 2 Gyr, highlighting the presence of evolved populations and the existence of Blue Straggler Star (BSS) members in this Galactic sector.

The \textit{ESFERA} sample reveals a clear evolutionary sequence of young stellar populations. \textit{ESFERA\,3} stands out as the youngest system in our census, with an age of 10 Myr and the presence of confirmed YSOs, indicating a very recent star-formation event. The rest of the young sample (with ages of 20, 30, and 50 Myr) completes this age distribution, providing a comprehensive snapshot of the star-formation history in this sector of the Carina–Sagittarius arm.

The implementation of the HDBSCAN clustering algorithm on {\it Gaia} DR3 astrometry proved to be a powerful tool for isolating these systems from the dense Galactic field. By combining this with a multi-wavelength validation (visual to radio), we were able to confirm the physical reality of these candidates, effectively overcoming the high extinction levels and stellar crowding characteristic of the Carina–Sagittarius sector. 
Our analysis of the interaction between the clusters and their environment reveals that spatial proximity does not necessarily imply a common origin. Indeed, stellar populations of different ages and evolutionary stages frequently overlap along the same line of sight within the spiral arm, highlighting the complex structural assembly of the Carina-Sagittarius region. This complexity underscores the need for a systematic census, which is continued in the second part of this study.

\begin{acknowledgements}
The authors thank the anonymous referee for the constructive comments and N. Duronea and C. Damia Rincón for useful feedback. This work used NASA's ADS, the SIMBAD database, and ALADIN tools (CDS, France). It also used 2MASS and WISE data (funded by NASA and NSF/UCLA/JPL), and the SUMSS radio survey (MOST, Australia). Data from the ESA mission {\it Gaia} (\url{https://www.cosmos.esa.int/gaia}) was processed by the {\it Gaia} DPAC (\url{https://www.cosmos.esa.int/web/gaia/dpac/consortium}).
\end{acknowledgements}
\bibliographystyle{aa}  
\bibliography{bibliografia}

\begin{thebibliography}{61}
\expandafter\ifx\csname natexlab\endcsname\relax\def\natexlab#1{#1}\fi

\bibitem[{{Allen} {et~al.}(2004){Allen}, {Calvet}, {D'Alessio}, {Mer{\'\i}n},
  {Hartmann}, {Megeath}, {Gutermuth}, {Muzerolle}, {Pipher}, {Myers}, \&
  {Fazio}}]{Allen04}
{Allen}, L.~E., {Calvet}, N., {D'Alessio}, P., {et~al.} 2004, \apjs, 154, 363

\bibitem[{{Anderson} {et~al.}(2014){Anderson}, {Bania}, {Balser}, {Cunningham},
  {Wenger}, {Johnstone}, \& {Armentrout}}]{Anderson14}
{Anderson}, L.~D., {Bania}, T.~M., {Balser}, D.~S., {et~al.} 2014, \apjs, 212,
  1

\bibitem[{{Ball} {et~al.}(2025){Ball}, {Kothes}, {Rosolowsky},
  {Burger-Scheidlin}, {Filipovi{\'c}}, {Lazarevi{\'c}}, {Smeaton}, {Becker},
  {Carretti}, {Gaensler}, {Hopkins}, {Leahy}, {Tahani}, {West}, {Anderson},
  {Loru}, {Ma}, {McClure-Griffiths}, \& {Micha{\l}owski}}]{Ball25}
{Ball}, B.~D., {Kothes}, R., {Rosolowsky}, E., {et~al.} 2025, \apj, 988, 75

\bibitem[{{Bock} {et~al.}(1999){Bock}, {Large}, \& {Sadler}}]{boc99}
{Bock}, D.~C.-J., {Large}, M.~I., \& {Sadler}, E.~M. 1999, \aj, 117, 1578

\bibitem[{{Borissova} {et~al.}(2012){Borissova}, {Georgiev}, {Hanson},
  {Clarke}, {Kurtev}, {Ivanov}, {Penaloza}, {Hillier}, \&
  {Zsarg{\'o}}}]{Borissova12}
{Borissova}, J., {Georgiev}, L., {Hanson}, M.~M., {et~al.} 2012, \aap, 546,
  A110

\bibitem[{Campello {et~al.}(2013)Campello, Moulavi, \& Sander}]{campello2013}
Campello, R. J. G.~B., Moulavi, D., \& Sander, J. 2013, in Advances in
  Knowledge Discovery and Data Mining, ed. J.~Pei, V.~S. Tseng, L.~Cao,
  H.~Motoda, \& G.~Xu (Berlin, Heidelberg: Springer Berlin Heidelberg),
  160--172

\bibitem[{{Cantat-Gaudin} \& {Anders}(2020)}]{Cantat2020Anders}
{Cantat-Gaudin}, T. \& {Anders}, F. 2020, \aap, 633, A99

\bibitem[{{Cantat-Gaudin} {et~al.}(2020){Cantat-Gaudin}, {Anders},
  {Castro-Ginard}, {Jordi}, {Romero-G{\'o}mez}, {Soubiran}, {Casamiquela},
  {Tarricq}, {Moitinho}, {Vallenari}, {Bragaglia}, {Krone-Martins}, \&
  {Kounkel}}]{2020cg}
{Cantat-Gaudin}, T., {Anders}, F., {Castro-Ginard}, A., {et~al.} 2020, A\&A,
  640, A1

\bibitem[{{Carpenter}(2001)}]{Carpenter01}
{Carpenter}, J.~M. 2001, \aj, 121, 2851

\bibitem[{{Case} \& {Bhattacharya}(1998)}]{Case98}
{Case}, G.~L. \& {Bhattacharya}, D. 1998, \apj, 504, 761

\bibitem[{{Castro-Ginard} {et~al.}(2024){Castro-Ginard}, {Penoyre}, {Casey},
  {Brown}, {Belokurov}, {Cantat-Gaudin}, {Drimmel}, {Fouesneau}, {Khanna},
  {Kurbatov}, {Price-Whelan}, {Rix}, \& {Smart}}]{2024cas}
{Castro-Ginard}, A., {Penoyre}, Z., {Casey}, A.~R., {et~al.} 2024, \aap, 688,
  A1

\bibitem[{{Cavallo} {et~al.}(2024){Cavallo}, {Spina}, {Carraro}, {Magrini},
  {Poggio}, {Cantat-Gaudin}, {Pasquato}, {Lucatello}, {Ortolani}, \&
  {Schiappacasse-Ulloa}}]{2024cavallo}
{Cavallo}, L., {Spina}, L., {Carraro}, G., {et~al.} 2024, \aj, 167, 12

\bibitem[{{Chen} {et~al.}(2019){Chen}, {Huang}, {Hou}, {Tian}, {Li}, {Yuan},
  {Wang}, {Wang}, {Tian}, \& {Liu}}]{2019Chen}
{Chen}, B.~Q., {Huang}, Y., {Hou}, L.~G., {et~al.} 2019, \mnras, 487, 1400

\bibitem[{{Choi} {et~al.}(2014){Choi}, {Hachisuka}, {Reid}, {Xu}, {Brunthaler},
  {Menten}, \& {Dame}}]{2014choi}
{Choi}, Y.~K., {Hachisuka}, K., {Reid}, M.~J., {et~al.} 2014, ApJ, 790, 99

\bibitem[{{Corti} {et~al.}(2023){Corti}, {Baume}, {Orellana}, \&
  {Suad}}]{Corti23}
{Corti}, M.~A., {Baume}, G.~L., {Orellana}, R.~B., \& {Suad}, L.~A. 2023, \aap,
  674, A55

\bibitem[{{Dean} {et~al.}(1978){Dean}, {Warren}, \& {Cousins}}]{Dean78}
{Dean}, J.~F., {Warren}, P.~R., \& {Cousins}, A.~W.~J. 1978, \mnras, 183, 569

\bibitem[{{Dias} {et~al.}(2021){Dias}, {Monteiro}, {Moitinho}, {L{\'e}pine},
  {Carraro}, {Paunzen}, {Alessi}, \& {Villela}}]{2021dias}
{Dias}, W.~S., {Monteiro}, H., {Moitinho}, A., {et~al.} 2021, MNRAS, 504, 356

\bibitem[{{Efremov}(1998)}]{efremov1998}
{Efremov}, Y.~N. 1998, Astronomical and Astrophysical Transactions, 15, 3

\bibitem[{{Efremov}(2010)}]{2010e}
{Efremov}, Y.~N. 2010, \mnras, 405, 1531

\bibitem[{{Elmegreen}(1982)}]{1982el}
{Elmegreen}, B.~G. 1982, \apj, 253, 655

\bibitem[{{Gaia Collaboration} {et~al.}(2016){Gaia Collaboration}, {Prusti},
  {de Bruijne}, {Brown}, {Vallenari}, {Babusiaux}, {Bailer-Jones}, {Bastian},
  {Biermann}, {Evans}, {Eyer}, {Jansen}, {Jordi}, {Klioner}, {Lammers},
  {Lindegren}, {Luri}, {Mignard}, {Milligan}, {Panem}, {Poinsignon},
  {Pourbaix}, {Randich}, {Sarri}, {Sartoretti}, {Siddiqui}, {Soubiran},
  {Valette}, {van Leeuwen}, {Walton}, {Aerts}, {Arenou}, {Cropper}, {Drimmel},
  {H{\o}g}, {Katz}, {Lattanzi}, {O'Mullane}, {Grebel}, {Holland}, {Huc},
  {Passot}, {Bramante}, {Cacciari}, {Casta{\~n}eda}, {Chaoul}, {Cheek}, {De
  Angeli}, {Fabricius}, {Guerra}, {Hern{\'a}ndez}, {Jean-Antoine-Piccolo},
  {Masana}, {Messineo}, {Mowlavi}, {Nienartowicz}, {Ord{\'o}{\~n}ez-Blanco},
  {Panuzzo}, {Portell}, {Richards}, {Riello}, {Seabroke}, {Tanga},
  {Th{\'e}venin}, {Torra}, {Els}, {Gracia-Abril}, {Comoretto},
  {Garcia-Reinaldos}, {Lock}, {Mercier}, {Altmann}, {Andrae}, {Astraatmadja},
  {Bellas-Velidis}, {Benson}, {Berthier}, {Blomme}, {Busso}, {Carry},
  {Cellino}, {Clementini}, {Cowell}, {Creevey}, {Cuypers}, {Davidson}, {De
  Ridder}, {de Torres}, {Delchambre}, {Dell'Oro}, {Ducourant}, {Fr{\'e}mat},
  {Garc{\'\i}a-Torres}, {Gosset}, {Halbwachs}, {Hambly}, {Harrison}, {Hauser},
  {Hestroffer}, {Hodgkin}, {Huckle}, {Hutton}, {Jasniewicz}, {Jordan},
  {Kontizas}, {Korn}, {Lanzafame}, {Manteiga}, {Moitinho}, {Muinonen},
  {Osinde}, {Pancino}, {Pauwels}, {Petit}, {Recio-Blanco}, {Robin}, {Sarro},
  {Siopis}, {Smith}, {Smith}, {Sozzetti}, {Thuillot}, {van Reeven}, {Viala},
  {Abbas}, {Abreu Aramburu}, {Accart}, {Aguado}, {Allan}, {Allasia},
  {Altavilla}, {{\'A}lvarez}, {Alves}, {Anderson}, {Andrei}, {Anglada Varela},
  {Antiche}, {Antoja}, {Ant{\'o}n}, {Arcay}, {Atzei}, {Ayache}, {Bach},
  {Baker}, {Balaguer-N{\'u}{\~n}ez}, {Barache}, {Barata}, {Barbier}, {Barblan},
  {Baroni}, {Barrado y Navascu{\'e}s}, {Barros}, {Barstow}, {Becciani},
  {Bellazzini}, {Bellei}, {Bello Garc{\'\i}a}, {Belokurov}, {Bendjoya},
  {Berihuete}, {Bianchi}, {Bienaym{\'e}}, {Billebaud}, {Blagorodnova},
  {Blanco-Cuaresma}, {Boch}, {Bombrun}, {Borrachero}, {Bouquillon}, {Bourda},
  {Bouy}, {Bragaglia}, {Breddels}, {Brouillet}, {Br{\"u}semeister},
  {Bucciarelli}, {Budnik}, {Burgess}, {Burgon}, {Burlacu}, {Busonero}, {Buzzi},
  {Caffau}, {Cambras}, {Campbell}, {Cancelliere}, {Cantat-Gaudin}, {Carlucci},
  {Carrasco}, {Castellani}, {Charlot}, {Charnas}, {Charvet}, {Chassat},
  {Chiavassa}, {Clotet}, {Cocozza}, {Collins}, {Collins}, \&
  {Costigan}}]{Gaia2016}
{Gaia Collaboration}, {Prusti}, T., {de Bruijne}, J.~H.~J., {et~al.} 2016,
  \aap, 595, A1

\bibitem[{{Gaia Collaboration} {et~al.}(2023){Gaia Collaboration}, {Vallenari},
  {Brown}, {Prusti}, {de Bruijne}, {Arenou}, {Babusiaux}, {Biermann},
  {Creevey}, {Ducourant}, {Evans}, {Eyer}, {Guerra}, {Hutton}, {Jordi},
  {Klioner}, {Lammers}, {Lindegren}, {Luri}, {Mignard}, {Panem}, {Pourbaix},
  {Randich}, {Sartoretti}, {Soubiran}, {Tanga}, {Walton}, {Bailer-Jones},
  {Bastian}, {Drimmel}, {Jansen}, {Katz}, {Lattanzi}, {van Leeuwen}, {Bakker},
  {Cacciari}, {Casta{\~n}eda}, {De Angeli}, {Fabricius}, {Fouesneau},
  {Fr{\'e}mat}, {Galluccio}, {Guerrier}, {Heiter}, {Masana}, {Messineo},
  {Mowlavi}, {Nicolas}, {Nienartowicz}, {Pailler}, {Panuzzo}, {Riclet}, {Roux},
  {Seabroke}, {Sordo}, {Th{\'e}venin}, {Gracia-Abril}, {Portell}, {Teyssier},
  {Altmann}, {Andrae}, {Audard}, {Bellas-Velidis}, {Benson}, {Berthier},
  {Blomme}, {Burgess}, {Busonero}, {Busso}, {C{\'a}novas}, {Carry}, {Cellino},
  {Cheek}, {Clementini}, {Damerdji}, {Davidson}, {de Teodoro}, {Nu{\~n}ez
  Campos}, {Delchambre}, {Dell'Oro}, {Esquej}, {Fern{\'a}ndez-Hern{\'a}ndez},
  {Fraile}, {Garabato}, {Garc{\'\i}a-Lario}, {Gosset}, {Haigron}, {Halbwachs},
  {Hambly}, {Harrison}, {Hern{\'a}ndez}, {Hestroffer}, {Hodgkin}, {Holl},
  {Jan{\ss}en}, {Jevardat de Fombelle}, {Jordan}, {Krone-Martins}, {Lanzafame},
  {L{\"o}ffler}, {Marchal}, {Marrese}, {Moitinho}, {Muinonen}, {Osborne},
  {Pancino}, {Pauwels}, {Recio-Blanco}, {Reyl{\'e}}, {Riello}, {Rimoldini},
  {Roegiers}, {Rybizki}, {Sarro}, {Siopis}, {Smith}, {Sozzetti}, {Utrilla},
  {van Leeuwen}, {Abbas}, {{\'A}brah{\'a}m}, {Abreu Aramburu}, {Aerts},
  {Aguado}, {Ajaj}, {Aldea-Montero}, {Altavilla}, {{\'A}lvarez}, {Alves},
  {Anders}, {Anderson}, {Anglada Varela}, {Antoja}, {Baines}, {Baker},
  {Balaguer-N{\'u}{\~n}ez}, {Balbinot}, {Balog}, {Barache}, {Barbato},
  {Barros}, {Barstow}, {Bartolom{\'e}}, {Bassilana}, {Bauchet}, {Becciani},
  {Bellazzini}, {Berihuete}, {Bernet}, {Bertone}, {Bianchi}, {Binnenfeld},
  {Blanco-Cuaresma}, {Blazere}, {Boch}, {Bombrun}, {Bossini}, {Bouquillon},
  {Bragaglia}, {Bramante}, {Breedt}, {Bressan}, {Brouillet}, {Brugaletta},
  {Bucciarelli}, {Burlacu}, {Butkevich}, {Buzzi}, {Caffau}, {Cancelliere},
  {Cantat-Gaudin}, {Carballo}, {Carlucci}, {Carnerero}, {Carrasco},
  {Casamiquela}, {Castellani}, {Castro-Ginard}, {Chaoul}, {Charlot}, {Chemin},
  {Chiaramida}, {Chiavassa}, {Chornay}, {Comoretto}, {Contursi}, {Cooper},
  {Cornez}, {Cowell}, {Crifo}, {Cropper}, {Crosta}, {Crowley}, {Dafonte},
  {Dapergolas}, {David}, {David}, {de Laverny}, {De Luise}, \& {De
  March}}]{Gaia2023}
{Gaia Collaboration}, {Vallenari}, A., {Brown}, A.~G.~A., {et~al.} 2023, \aap,
  674, A1

\bibitem[{{Gooch}(1996)}]{Gooch96}
{Gooch}, R. 1996, in Astronomical Society of the Pacific Conference Series,
  Vol. 101, Astronomical Data Analysis Software and Systems V, ed. G.~H.
  {Jacoby} \& J.~{Barnes}, 80

\bibitem[{{Green}(2019)}]{Green19}
{Green}, D.~A. 2019, Journal of Astrophysics and Astronomy, 40, 36

\bibitem[{{Hao} {et~al.}(2022){Hao}, {Xu}, {Wu}, {Lin}, {Liu}, \&
  {Li}}]{2022hao}
{Hao}, C.~J., {Xu}, Y., {Wu}, Z.~Y., {et~al.} 2022, \aap, 660, A4

\bibitem[{{Hou} \& {Han}(2014)}]{HouHan14}
{Hou}, L.~G. \& {Han}, J.~L. 2014, \aap, 569, A125

\bibitem[{{Hunt} {et~al.}(2025){Hunt}, {Cantat-Gaudin}, {Anders}, {Spina},
  {Cavallo}, {Castro-Ginard}, {Belokurov}, {Brown}, {Casey}, {Drimmel},
  {Fouesneau}, \& {Reffert}}]{Hunt2025}
{Hunt}, E.~L., {Cantat-Gaudin}, T., {Anders}, F., {et~al.} 2025, \aap, 699,
  A273

\bibitem[{{Hunt} \& {Reffert}(2021)}]{Hunt2021}
{Hunt}, E.~L. \& {Reffert}, S. 2021, \aap, 646, A104

\bibitem[{{Hunt} \& {Reffert}(2023)}]{2023hunt}
{Hunt}, E.~L. \& {Reffert}, S. 2023, \aap, 673, A114

\bibitem[{{Janes} \& {Phelps}(1994)}]{Janes94}
{Janes}, K.~A. \& {Phelps}, R.~L. 1994, \aj, 108, 1773

\bibitem[{{Jordi} {et~al.}(2010){Jordi}, {Gebran}, {Carrasco}, {de Bruijne},
  {Voss}, {Fabricius}, {Knude}, {Vallenari}, {Kohley}, \& {Mora}}]{Jordi10}
{Jordi}, C., {Gebran}, M., {Carrasco}, J.~M., {et~al.} 2010, \aap, 523, A48

\bibitem[{{Joshi}(2025)}]{Joshi25}
{Joshi}, Y.~C. 2025, \na, 121, 102425

\bibitem[{{Joye} {et~al.}(2025){Joye}, {oldherl}, {Burke}, \&
  {Glotfelty}}]{Joye25}
{Joye}, W., {oldherl}, {Burke}, D., \& {Glotfelty}, K. 2025,
  {SAOImageDS9/SAOImageDS9: v8.7b1}

\bibitem[{{Kim} \& {Ostriker}(2002)}]{2002kim}
{Kim}, W.-T. \& {Ostriker}, E.~C. 2002, \apj, 570, 132

\bibitem[{{Koenig} \& {Leisawitz}(2014)}]{Koenig14}
{Koenig}, X.~P. \& {Leisawitz}, D.~T. 2014, \apj, 791, 131

\bibitem[{{Koornneef}(1983)}]{Koornneef83}
{Koornneef}, J. 1983, \aap, 128, 84

\bibitem[{{Lada} \& {Adams}(1992)}]{Lada92}
{Lada}, C.~J. \& {Adams}, F.~C. 1992, \apj, 393, 278

\bibitem[{{Landolt-Bornstein} \& {Bursa}(1982)}]{LB82}
{Landolt-Bornstein} \& {Bursa}, M. 1982, Bulletin of the Astronomical
  Institutes of Czechoslovakia, 33, 380

\bibitem[{{Liu} {et~al.}(2025){Liu}, {He}, {Luo}, \& {Wang}}]{2025liu}
{Liu}, X., {He}, Z., {Luo}, Y., \& {Wang}, K. 2025, \mnras, 537, 2403

\bibitem[{{Messineo} {et~al.}(2012){Messineo}, {Menten}, {Churchwell}, \&
  {Habing}}]{Messineo12}
{Messineo}, M., {Menten}, K.~M., {Churchwell}, E., \& {Habing}, H. 2012, \aap,
  537, A10

\bibitem[{{Molina Lera} {et~al.}(2018){Molina Lera}, {Baume}, \&
  {Gamen}}]{Molina2018}
{Molina Lera}, J.~A., {Baume}, G., \& {Gamen}, R. 2018, MNRAS, 480, 2386

\bibitem[{{Murray} \& {Rahman}(2010)}]{Murray10}
{Murray}, N. \& {Rahman}, M. 2010, \apj, 709, 424

\bibitem[{{Negueruela} {et~al.}(2007){Negueruela}, {Marco}, {Israel}, \&
  {Bernabeu}}]{Negueruela07}
{Negueruela}, I., {Marco}, A., {Israel}, G.~L., \& {Bernabeu}, G. 2007, \aap,
  471, 485

\bibitem[{{Pa{\'\i}z} {et~al.}(2025){Pa{\'\i}z}, {Biasi}, {Orellana}, \&
  {Corti}}]{Paiz2025}
{Pa{\'\i}z}, L.~G., {Biasi}, M. S.~D., {Orellana}, R.~B., \& {Corti}, M.~A.
  2025, International Journal of Astronomy and Astrophysics, 15, 171

\bibitem[{{Pedregosa} {et~al.}(2011){Pedregosa}, {Varoquaux}, {Gramfort},
  {Michel}, {Thirion}, {Grisel}, {Blondel}, {M{\"u}ller}, {Nothman}, {Louppe},
  {Prettenhofer}, {Weiss}, {Dubourg}, {Vanderplas}, {Passos}, {Cournapeau},
  {Brucher}, {Perrot}, \& {Duchesnay}}]{pedregosa2011}
{Pedregosa}, F., {Varoquaux}, G., {Gramfort}, A., {et~al.} 2011, Journal of
  Machine Learning Research, 12, 2825

\bibitem[{{Penoyre} {et~al.}(2022){Penoyre}, {Belokurov}, \&
  {Evans}}]{2022Penoyre}
{Penoyre}, Z., {Belokurov}, V., \& {Evans}, N.~W. 2022, \mnras, 513, 2437

\bibitem[{{Perren} {et~al.}(2023){Perren}, {Pera}, {Navone}, \&
  {V{\'a}zquez}}]{Perren23}
{Perren}, G.~I., {Pera}, M.~S., {Navone}, H.~D., \& {V{\'a}zquez}, R.~A. 2023,
  \mnras, 526, 4107

\bibitem[{{Poleski}(2013)}]{Poleski2013}
{Poleski}, R. 2013, arXiv e-prints, arXiv:1306.2945

\bibitem[{{Rain} {et~al.}(2024){Rain}, {Pera}, {Perren}, {Benvenuto}, {Panei},
  {De Vito}, {Carraro}, \& {Villanova}}]{Rain24}
{Rain}, M.~J., {Pera}, M.~S., {Perren}, G.~I., {et~al.} 2024, \aap, 685, A33

\bibitem[{{Rieke} \& {Lebofsky}(1985)}]{Rieke85}
{Rieke}, G.~H. \& {Lebofsky}, M.~J. 1985, \apj, 288, 618

\bibitem[{{Rizzo} {et~al.}(2025){Rizzo}, {Corti}, \& {Pa{\'\i}z}}]{2025esfera}
{Rizzo}, L., {Corti}, M.~A., \& {Pa{\'\i}z}, L.~G. 2025, Boletin de la
  Asociacion Argentina de Astronomia La Plata Argentina, 66, 181

\bibitem[{{Santos-Silva} {et~al.}(2021){Santos-Silva}, {Perottoni},
  {Almeida-Fernandes}, {Gregorio-Hetem}, {Jatenco-Pereira}, {Mendes de
  Oliveira}, {Montmerle}, {Bica}, {Bonatto}, {Monteiro}, {Dias}, {Barbosa},
  {Fernandes}, {Galli}, {Borges Fernandes}, {Kanaan}, {Ribeiro}, \&
  {Schoenell}}]{SantosSilva2021}
{Santos-Silva}, T., {Perottoni}, H.~D., {Almeida-Fernandes}, F., {et~al.} 2021,
  \mnras, 508, 1033

\bibitem[{{Skrutskie} {et~al.}(2006){Skrutskie}, {Cutri}, {Stiening},
  {Weinberg}, {Schneider}, {Carpenter}, {Beichman}, {Capps}, {Chester},
  {Elias}, {Huchra}, {Liebert}, {Lonsdale}, {Monet}, {Price}, {Seitzer},
  {Jarrett}, {Kirkpatrick}, {Gizis}, {Howard}, {Evans}, {Fowler}, {Fullmer},
  {Hurt}, {Light}, {Kopan}, {Marsh}, {McCallon}, {Tam}, {Van Dyk}, \&
  {Wheelock}}]{Skrutskie06}
{Skrutskie}, M.~F., {Cutri}, R.~M., {Stiening}, R., {et~al.} 2006, \aj, 131,
  1163

\bibitem[{{Spitzer}(1958)}]{Spitzer58}
{Spitzer}, Jr., L. 1958, \apj, 127, 17

\bibitem[{{Taylor}(2011)}]{Taylor11}
{Taylor}, M. 2011, {TOPCAT: Tool for OPerations on Catalogues And Tables},
  Astrophysics Source Code Library, record ascl:1101.010

\bibitem[{{V{\'a}zquez} {et~al.}(2008){V{\'a}zquez}, {May}, {Carraro},
  {Bronfman}, {Moitinho}, \& {Baume}}]{2008va}
{V{\'a}zquez}, R.~A., {May}, J., {Carraro}, G., {et~al.} 2008, ApJ, 672, 930

\bibitem[{{Wang} \& {Chen}(2019)}]{Wang19}
{Wang}, S. \& {Chen}, X. 2019, \apj, 877, 116

\bibitem[{{Wielen}(1985)}]{Wielen85}
{Wielen}, R. 1985, in IAU Symposium, Vol. 113, Dynamics of Star Clusters, ed.
  J.~{Goodman} \& P.~{Hut}, 449--460

\bibitem[{{Wienen} {et~al.}(2015){Wienen}, {Wyrowski}, {Menten}, {Urquhart},
  {Walmsley}, {Csengeri}, {Koribalski}, \& {Schuller}}]{Wienen15}
{Wienen}, M., {Wyrowski}, F., {Menten}, K.~M., {et~al.} 2015, \aap, 579, A91

\bibitem[{{Wright} {et~al.}(2010){Wright}, {Eisenhardt}, {Mainzer}, {Ressler},
  {Cutri}, {Jarrett}, {Kirkpatrick}, {Padgett}, {McMillan}, {Skrutskie},
  {Stanford}, {Cohen}, {Walker}, {Mather}, {Leisawitz}, {Gautier}, {McLean},
  {Benford}, {Lonsdale}, {Blain}, {Mendez}, {Irace}, {Duval}, {Liu}, {Royer},
  {Heinrichsen}, {Howard}, {Shannon}, {Kendall}, {Walsh}, {Larsen}, {Cardon},
  {Schick}, {Schwalm}, {Abid}, {Fabinsky}, {Naes}, \& {Tsai}}]{Wri10}
{Wright}, E.~L., {Eisenhardt}, P. R.~M., {Mainzer}, A.~K., {et~al.} 2010, \aj,
  140, 1868

\bibitem[{{Zorec} \& {Briot}(1997)}]{Zorec97}
{Zorec}, J. \& {Briot}, D. 1997, \aap, 318, 443

\end{thebibliography}

\begin{appendix}

\section{Astrometric results}

\begin{table*}
\caption{Probable members of open cluster $ESFERA\,1$ from astrometric analysis}
\label{table:astroesf1}
\centering
\fontsize{9pt}{10pt}\selectfont
\setlength{\tabcolsep}{3pt}
\begin{tabular}{c c c c c c c c c} 
    \toprule
    $GAIA-DR3$ & $\alpha$ & $\delta$ & $\varpi$ & $\epsilon_{\varpi}$ & $\mu_{\alpha}\rm{cos}\delta$ & $\epsilon_{\mu_{\alpha}\rm{cos}\delta}$ & $\mu_{\delta}$ & $\epsilon_{\mu_{\delta}}$ \\
    \midrule 
    58800...& $[\circ:':'']$ & $[\circ:':'']$ & $[mas]$ & $[mas]$ & [mas yr$^{-1}$] &[mas yr$^{-1}$] & [mas yr$^{-1}$] &[mas yr$^{-1}$]  \\ 
    \midrule 
    \begin{filecontents*}{AstromEsfera1.csv}
GAIA,AR,DEC,PARAL,ePARAL,PMAR,ePMAR,PMDEC,ePMDEC
26973766044928,227:11:07.0,-58:31:28.6,0.317,0.053,-3.618,0.052,-2.968,0.072
27798399504896,227:25:12.3,-58:29:59.6,0.507,0.060,-3.267,0.061,-2.860,0.070
29378947544832,227:21:49.2,-58:24:55.1,0.374,0.029,-3.622,0.029,-3.054,0.039
29413307278848,227:22:33.7,-58:25:07.0,0.460,0.078,-3.495,0.081,-2.802,0.101
29516386509824,227:23:16.1,-58:23:43.6,0.421,0.057,-3.482,0.056,-2.948,0.064
29619465728256,227:21:02.2,-58:23:54.4,0.418,0.020,-3.329,0.019,-2.827,0.026
29619465729152,227:21:13.9,-58:23:46.0,0.412,0.063,-3.271,0.064,-2.797,0.083
29619470340480,227:21:11.4,-58:23:33.2,0.434,0.080,-3.398,0.078,-2.796,0.103
29756904691200,227:22:44.9,-58:22:32.0,0.408,0.053,-3.271,0.050,-2.842,0.062
29791264429696,227:21:10.7,-58:22:51.2,0.391,0.072,-3.454,0.072,-2.855,0.094
38862235534720,227:07:24.1,-58:29:33.5,0.372,0.047,-3.946,0.049,-3.118,0.057
39205832933632,227:10:40.6,-58:28:41.1,0.382,0.027,-3.546,0.027,-2.859,0.036
39549430334336,227:07:59.2,-58:26:46.0,0.375,0.030,-3.436,0.030,-2.838,0.040
40855100166784,227:15:02.6,-58:25:51.7,0.356,0.032,-3.400,0.033,-2.776,0.041
52193818076672,227:26:41.3,-58:23:43.2,0.348,0.054,-3.453,0.052,-2.724,0.063
52262534093312,227:30:39.2,-58:23:23.5,0.387,0.050,-3.362,0.050,-2.752,0.056
\end{filecontents*}
\csvreader[
        head to column names, 
        separator=comma,
        late after line=\\
    ]{AstromEsfera1.csv}{}{
        \GAIA & \AR & \DEC & \PARAL & \ePARAL & \PMAR &  \ePMAR & \PMDEC & \ePMDEC  
    }
    \bottomrule 
\end{tabular}
\end{table*}                                  

\begin{table*}
\caption{Probable members of open cluster $ESFERA\,2$ from astrometric analysis}
\label{table:astroesf2}
\centering
\fontsize{9pt}{10pt}\selectfont
\setlength{\tabcolsep}{3pt}
\begin{tabular}{c c c c c c c c c} 
    \toprule
    $GAIA-DR3$ & $\alpha$ & $\delta$ & $\varpi$ & $\epsilon_{\varpi}$ & $\mu_{\alpha}\rm{cos}\delta$ & $\epsilon_{\mu_{\alpha}\rm{cos}\delta}$ & $\mu_{\delta}$ & $\epsilon_{\mu_{\delta}}$ \\
    \midrule 
    58831...& $[\circ:':'']$ & $[\circ:':'']$ & $[mas]$ & $[mas]$ & [mas yr$^{-1}$] &[mas yr$^{-1}$] & [mas yr$^{-1}$] &[mas yr$^{-1}$]  \\ 
    \midrule 
    \begin{filecontents*}{AstromEsfera2.csv}
GAIA,AR,DEC,PARAL,ePARAL,PMAR,ePMAR,PMDEC,ePMDEC
17048125978624,230:09:20.4,-57:49:33.5,0.551,0.087,-3.618,0.089,-3.359,0.084
22923616738560,230:16:45.3,-57:47:41.8,0.513,0.057,-3.631,0.057,-3.529,0.055
23022399628544,230:16:02.0,-57:45:33.7,0.498,0.096,-3.273,0.094,-3.428,0.092
23335958552320,230:18:32.2,-57:45:50.0,0.481,0.069,-3.656,0.068,-3.447,0.066
23439037784960,230:17:06.3,-57:44:46.4,0.453,0.023,-3.655,0.023,-3.470,0.023
23507732445696,230:20:19.0,-57:43:25.0,0.533,0.040,-3.592,0.041,-3.453,0.040
23576476750208,230:19:04.8,-57:43:31.2,0.566,0.045,-3.398,0.045,-3.540,0.046
23679555965184,230:14:50.9,-57:44:31.9,0.553,0.023,-3.696,0.022,-3.460,0.025
23748275451136,230:17:38.8,-57:43:15.3,0.548,0.045,-3.582,0.044,-3.528,0.043
23748275452288,230:16:56.7,-57:43:19.1,0.534,0.022,-3.659,0.021,-3.294,0.022
24328068894592,230:25:51.3,-57:42:41.4,0.508,0.072,-3.592,0.074,-3.435,0.068
25500622168576,230:20:19.3,-57:38:44.1,0.528,0.089,-3.703,0.086,-3.801,0.087
27935846172416,229:55:34.4,-57:49:35.4,0.517,0.087,-3.711,0.086,-3.644,0.079
29658150063488,229:56:53.0,-57:47:22.0,0.466,0.026,-3.294,0.025,-3.627,0.025
33128483745408,229:55:00.3,-57:43:01.6,0.487,0.020,-3.446,0.019,-3.378,0.020
36117781323776,230:07:19.0,-57:41:46.5,0.521,0.074,-3.769,0.070,-3.663,0.075
37590927975040,230:16:37.4,-57:36:35.9,0.482,0.080,-3.762,0.075,-3.439,0.080
\end{filecontents*}
\csvreader[
        head to column names, 
        separator=comma,
        late after line=\\
    ]{AstromEsfera2.csv}{}{
        \GAIA & \AR & \DEC & \PARAL & \ePARAL & \PMAR &  \ePMAR & \PMDEC & \ePMDEC  
    }
    \bottomrule 
\end{tabular}
\end{table*}                                  

\begin{table*}
\caption{Probable members of open cluster $ESFERA\,3$ from astrometric analysis}
\label{table:astroesf3}
\centering
\fontsize{9pt}{10pt}\selectfont
\setlength{\tabcolsep}{3pt}
\begin{tabular}{c c c c c c c c c} 
    \toprule
    $GAIA-DR3$ & $\alpha$ & $\delta$ & $\varpi$ & $\epsilon_{\varpi}$ & $\mu_{\alpha}\rm{cos}\delta$ & $\epsilon_{\mu_{\alpha}\rm{cos}\delta}$ & $\mu_{\delta}$ & $\epsilon_{\mu_{\delta}}$ \\
    \midrule 
    5883...& $[\circ:':'']$ & $[\circ:':'']$ & $[mas]$ & $[mas]$ & [mas yr$^{-1}$] &[mas yr$^{-1}$] & [mas yr$^{-1}$] &[mas yr$^{-1}$]  \\ 
    \midrule 
    \begin{filecontents*}{AstromEsfera3.csv}
GAIA,AR,DEC,PARAL,ePARAL,PMAR,ePMAR,PMDEC,ePMDEC
194941656356480,229:56:14.8,-57:18:33.4,0.396,0.044,-2.368,0.045,-2.173,0.045
214526707667072,230:05:26.7,-57:19:29.5,0.429,0.049,-2.309,0.052,-2.391,0.050
214973385226112,230:08:17.9,-57:17:50.0,0.581,0.042,-2.546,0.041,-2.152,0.044
215003418857472,230:09:51.0,-57:17:17.6,0.530,0.033,-2.345,0.034,-2.224,0.035
216210335393536,230:05:36.8,-57:13:30.6,0.406,0.055,-2.421,0.059,-2.212,0.061
216622652247808,230:11:22.8,-57:14:40.3,0.517,0.063,-1.925,0.067,-2.377,0.068
216863170424576,230:16:16.2,-57:14:10.7,0.522,0.019,-2.319,0.020,-1.854,0.021
217172408108160,230:14:01.6,-57:11:47.2,0.480,0.027,-2.142,0.026,-2.065,0.030
217172408111232,230:14:16.3,-57:11:40.4,0.528,0.040,-2.202,0.041,-1.756,0.044
217412926228736,230:08:15.5,-57:14:34.2,0.506,0.065,-1.930,0.068,-2.269,0.070
218031401599744,230:13:51.9,-57:10:21.6,0.575,0.039,-2.213,0.037,-2.019,0.044
219057869997952,229:59:54.0,-57:13:47.9,0.573,0.106,-2.072,0.111,-2.254,0.109
219440150823424,230:02:08.3,-57:11:45.4,0.451,0.046,-2.355,0.040,-2.235,0.047
221020698860928,230:02:25.8,-57:08:29.2,0.458,0.070,-2.260,0.063,-1.963,0.077
223151002640256,230:18:06.9,-57:09:42.9,0.590,0.054,-2.430,0.052,-2.359,0.063
232221972981504,229:54:05.5,-57:08:47.5,0.426,0.057,-2.391,0.052,-2.174,0.060
232359411945472,229:55:15.3,-57:07:05.9,0.493,0.040,-2.066,0.037,-1.919,0.044
234524076184960,230:01:42.8,-57:01:24.5,0.459,0.052,-1.656,0.052,-2.060,0.060
235344386772224,229:51:56.2,-57:00:58.9,0.457,0.067,-2.140,0.069,-2.335,0.078
236826178920448,229:46:16.3,-57:00:58.9,0.472,0.028,-2.263,0.028,-2.270,0.033
\end{filecontents*}
\csvreader[
        head to column names, 
        separator=comma,
        late after line=\\
    ]{AstromEsfera3.csv}{}{
        \GAIA & \AR & \DEC & \PARAL & \ePARAL & \PMAR &  \ePMAR & \PMDEC & \ePMDEC  
    }
    \bottomrule 
\end{tabular}
\end{table*}                                  

\begin{table*}
\caption{Probable members of open cluster $ESFERA\,4$ from astrometric analysis}
\label{table:astroesf4}
\centering
\fontsize{9pt}{10pt}\selectfont
\setlength{\tabcolsep}{3pt}
\begin{tabular}{c c c c c c c c c} 
    \toprule
    $GAIA-DR3$ & $\alpha$ & $\delta$ & $\varpi$ & $\epsilon_{\varpi}$ & $\mu_{\alpha}\rm{cos}\delta$ & $\epsilon_{\mu_{\alpha}\rm{cos}\delta}$ & $\mu_{\delta}$ & $\epsilon_{\mu_{\delta}}$ \\
    \midrule 
    5886...& $[\circ:':'']$ & $[\circ:':'']$ & $[mas]$ & $[mas]$ & [mas yr$^{-1}$] &[mas yr$^{-1}$] & [mas yr$^{-1}$] &[mas yr$^{-1}$]  \\ 
    \midrule 
    \begin{filecontents*}{AstromEsfera4.csv}
GAIA,AR,DEC,PARAL,ePARAL,PMAR,ePMAR,PMDEC,ePMDEC
257425123101568,229:43:21.2,-56:28:16.7,0.586,0.076,-2.476,0.086,-2.956,0.091
348100474124800,230:06:36.5,-56:28:09.0,0.489,0.095,-2.181,0.093,-2.990,0.105
349440504533248,230:07:29.8,-56:24:58.2,0.550,0.051,-2.363,0.054,-3.003,0.062
350643094839552,229:55:46.6,-56:26:12.8,0.593,0.045,-2.388,0.049,-3.067,0.056
351364649407360,229:46:52.6,-56:25:31.7,0.653,0.020,-2.545,0.023,-3.099,0.024
351429059257984,229:46:50.1,-56:25:22.8,0.576,0.014,-2.519,0.015,-2.927,0.017
352429801298816,229:56:06.5,-56:23:14.1,0.569,0.063,-2.622,0.063,-2.946,0.065
352567240269312,229:57:45.4,-56:21:57.5,0.481,0.068,-2.243,0.068,-3.093,0.070
353220075304320,229:52:54.0,-56:22:41.4,0.618,0.056,-2.451,0.060,-2.964,0.065
353769831159936,229:56:37.7,-56:19:46.8,0.500,0.048,-2.366,0.046,-2.971,0.051
363287478678016,229:44:56.3,-56:23:02.1,0.577,0.061,-2.364,0.061,-3.460,0.068
363386248420736,229:40:06.7,-56:24:05.8,0.593,0.094,-2.444,0.103,-2.627,0.115
363871594262144,229:45:03.2,-56:21:38.7,0.589,0.023,-2.653,0.022,-3.169,0.027
364318270911360,229:41:39.7,-56:20:32.6,0.596,0.038,-2.374,0.037,-2.841,0.049
\end{filecontents*}
\csvreader[
        head to column names, 
        separator=comma,
        late after line=\\
    ]{AstromEsfera4.csv}{}{
        \GAIA & \AR & \DEC & \PARAL & \ePARAL & \PMAR &  \ePMAR & \PMDEC & \ePMDEC  
    }
    \bottomrule 
\end{tabular}
\end{table*}                                  

\begin{table*}
\caption{Probable members of open cluster $ESFERA\,5$ from astrometric analysis}
\label{table:astroesf5}
\centering
\fontsize{9pt}{10pt}\selectfont
\setlength{\tabcolsep}{3pt}
\begin{tabular}{c c c c c c c c c} 
    \toprule
    $GAIA-DR3$ & $\alpha$ & $\delta$ & $\varpi$ & $\epsilon_{\varpi}$ & $\mu_{\alpha}\rm{cos}\delta$ & $\epsilon_{\mu_{\alpha}\rm{cos}\delta}$ & $\mu_{\delta}$ & $\epsilon_{\mu_{\delta}}$ \\
    \midrule 
    58838...& $[\circ:':'']$ & $[\circ:':'']$ & $[mas]$ & $[mas]$ & [mas yr$^{-1}$] &[mas yr$^{-1}$] & [mas yr$^{-1}$] &[mas yr$^{-1}$]  \\ 
    \midrule 
     \begin{filecontents*}{AstromEsfera5.csv}
GAIA,AR,DEC,PARAL,ePARAL,PMAR,ePMAR,PMDEC,ePMDEC
37086485049088,232:35:41.2,-55:23:40.5,0.355,0.031,-3.329,0.029,-2.987,0.031
37915429150592,232:25:37.4,-55:22:08.2,0.397,0.077,-3.263,0.073,-3.182,0.076
38911861950720,232:40:44.8,-55:26:33.2,0.431,0.048,-3.077,0.047,-3.197,0.050
40698568446208,232:40:51.0,-55:19:25.7,0.416,0.069,-3.300,0.064,-3.341,0.070
40732928203776,232:41:47.2,-55:18:14.3,0.364,0.066,-3.189,0.062,-3.010,0.067
41832439453056,232:52:12.3,-55:15:44.0,0.394,0.067,-3.286,0.071,-3.130,0.073
42210396958720,232:43:22.6,-55:17:55.6,0.383,0.030,-3.293,0.029,-3.055,0.033
43992792751232,232:38:00.4,-55:16:36.6,0.363,0.041,-3.174,0.039,-3.449,0.040
44100182546944,232:32:14.5,-55:18:46.8,0.372,0.055,-3.781,0.051,-3.101,0.057
44375059980800,232:30:16.1,-55:17:22.2,0.373,0.044,-3.789,0.042,-3.540,0.045
44409419719040,232:28:49.0,-55:17:50.0,0.408,0.050,-3.204,0.047,-3.113,0.051
44512499441280,232:31:34.4,-55:16:21.4,0.412,0.014,-3.297,0.014,-3.066,0.016
44649938394624,232:36:00.1,-55:16:36.8,0.415,0.024,-3.369,0.022,-3.083,0.024
44993535795840,232:32:12.7,-55:14:51.8,0.391,0.032,-2.753,0.030,-2.795,0.033
45165334464256,232:41:27.4,-55:17:29.1,0.427,0.049,-2.633,0.046,-3.084,0.050
91825860877184,232:25:42.7,-55:13:49.0,0.369,0.033,-3.575,0.034,-3.192,0.033
\end{filecontents*}
\csvreader[
        head to column names, 
        separator=comma,
        late after line=\\
    ]{AstromEsfera5.csv}{}{
           \GAIA & \AR & \DEC & \PARAL & \ePARAL & \PMAR &  \ePMAR & \PMDEC & \ePMDEC  
    }
    \bottomrule 
\end{tabular}
\end{table*}    

\section {Photometric results}

\begin{table*}
\caption{Probable members of open cluster $ESFERA\,1$ from photometric observations}
\label{table:esfera1p}
\centering
\fontsize{9pt}{10pt}\selectfont
\setlength{\tabcolsep}{3pt}
\begin{tabular}{c c c c c c r r r l} 
    \toprule
    $GAIA-DR3$ & $GLON$ & $GLAT$ & $G$ & $G_{BP}-G_{RP}$ & $J$ & $H$ & $K_s$ & $Q_{NIR}$ & $SpT$ \\
    \midrule 
     58800...& $[{\circ}:':'']$ & $[{\circ}:':'']$ & $[mag]$ & $[mag]$ & $[mag]$ & $[mag]$ & $[mag]$& & \\ 
    \midrule 
    \begin{filecontents*}{Esfera1P.csv}
GAIA,GLON,GLAT,Gmag,BPRP,Jmag,Hmag,Kmag,QNIR,TE
26973766044928,320:03:19.7,-00:17:38.1,16.17,1.75,13.84,13.36,13.31,0.38,B$^+$
27798399504896,320:10:25.9,-00:20:03.4,16.50,1.78,14.04,13.51,13.29,0.16,A$^-$
29378947544832,320:11:27.9,-00:14:46.8,15.24,1.58,13.10,12.68,12.51,0.12,B$^+$
29413307278848,320:11:42.0,-00:15:08.9,17.11,1.78,14.72,14.31,14.20,0.22,A$^+$
29516386509824,320:12:43.3,-00:14:08.1,16.50,1.75,14.15,13.75,13.63,0.18,A$^-$
29619465728256,320:11:37.2,-00:13:42.0,14.12,1.67,11.89,11.50,11.31,0.07,B$^+$
29619465729152,320:11:46.7,-00:13:37.9,16.77,1.81,14.32,13.92,13.70,0.01,A$^-$
29619470340480,320:11:52.1,-00:13:26.2,16.99,1.91,14.53,14.06,13.97,0.31,A$^+$
29756904691200,320:13:05.2,-00:12:58.1,16.41,1.85,13.96,13.54,13.33,0.05,B$^+$
29791264429696,320:12:12.9,-00:12:49.7,16.98,1.92,14.36,13.87,13.60,0.03,A$^-$
38862235534720,320:02:36.7,-00:15:00.1,15.90,3.03,12.18,11.18,10.76,0.30,B$^+$
39205832933632,320:04:31.8,-00:15:06.2,14.16,3.26,10.27,9.10,8.68,0.45,B$^+$
39549430334336,320:04:16.5,-00:12:44.3,15.23,1.83,12.82,12.44,12.16,-0.09,B$^+$
40855100166784,320:07:55.4,-00:13:48.6,15.31,1.83,12.94,12.54,12.32,0.03,B$^+$
52193818076672,320:14:16.3,-00:15:02.0,16.44,1.85,14.01,13.54,13.35,0.15,B$^+$
52262534093312,320:16:13.8,-00:15:48.1,16.32,1.89,13.77,13.35,13.12,0.03,B$^+$
\end{filecontents*}
\csvreader[
        head to column names, 
        separator=comma,
        late after line=\\
    ]{Esfera1P.csv}{}{
            \GAIA & \GLON & \GLAT & \Gmag & \BPRP & \Jmag &  \Hmag & \Kmag & \QNIR & \TE 
    }
    \bottomrule 
\end{tabular}
\end{table*}

\begin{table*}
\caption{Probable members of open cluster $ESFERA\,2$ from photometric observations}
\label{table:esfera2p}
\centering
\fontsize{9pt}{10pt}\selectfont
\setlength{\tabcolsep}{3pt}
\begin{tabular}{c c c c c r r r r l} 
    \toprule
    $GAIA-DR3$ & $GLON$ & $GLAT$ & $G$ & $G_{BP}-G_{RP}$ & $J$ & $H$ & $K_s$ & $Q_{NIR}$ & $SpT$ \\
    \midrule 
    58831...& $[{\circ}:':'']$ & $[{\circ}:':'']$ & $[mag]$ & $[mag]$ & $[mag]$ & $[mag]$ & $[mag]$& & \\ 
    \midrule 
    \begin{filecontents*}{Esfera2P.csv}
GAIA,GLON,GLAT,Gmag,BPRP,Jmag,Hmag,Kmag,QNIR,TE
17048125978624,321:45:22.8,-00:30:47.5,17.35,2.25,14.48,13.89,13.69,0.26,A$^+$
22923616738560,321:49:42.6,-00:31:21.5,16.58,1.87,14.11,13.57,13.35,0.16,A$^+$
23022399628544,321:50:32.5,-00:29:12.2,17.57,2.49,14.49,14.02,13.76,0.02,A$^+$
23335958552320,321:51:31.0,-00:30:01.8,16.97,1.98,14.32,13.82,13.55,0.05,F$^-$
23439037784960,321:51:26.9,-00:29:00.0,14.74,1.53,12.69,12.37,12.21,0.04,B$^+$
23507732445696,321:53:37.5,-00:28:47.3,15.83,1.77,13.51,13.06,12.89,0.17,A$^-$
23576476750208,321:53:07.9,-00:28:31.8,16.15,1.76,13.79,13.43,13.26,0.07,A$^+$
23679555965184,321:50:33.9,-00:28:08.8,10.24,2.71,6.85,5.92,5.56,0.32,CLIII-I
23748275451136,321:52:30.8,-00:27:52.7,16.12,1.98,13.58,13.07,12.88,0.18,A$^-$
23748275452288,321:52:09.8,-00:27:43.8,14.47,1.61,12.32,11.94,11.79,0.11,B$^+$
24328068894592,321:56:30.3,-00:29:46.8,17.05,2.53,13.74,13.31,13.08,0.03,B$^+$
25500622168576,321:56:09.8,-00:24:51.5,17.41,2.12,14.64,13.97,13.67,0.16,F$^-$
27935846172416,321:39:11.0,-00:26:52.5,17.09,2.65,13.71,12.92,12.31,-0.23,B$^+$
29658150063488,321:40:57.9,-00:25:22.4,14.93,1.63,12.72,12.39,12.22,0.02,B$^+$
33128483745408,321:42:26.9,-00:21:10.4,14.37,2.43,11.25,10.49,10.22,0.31,B$^+$
36117781323776,321:48:40.1,-00:23:39.2,16.89,1.94,14.44,14.04,13.75,-0.09,F$^-$
37590927975040,321:55:39.2,-00:21:59.1,17.22,1.86,--,--,--,--,F$^-$
\end{filecontents*}
\csvreader[
        head to column names, 
        separator=comma,
        late after line=\\
    ]{Esfera2P.csv}{}{
        \GAIA & \GLON & \GLAT & \Gmag & \BPRP & \Jmag &  \Hmag & \Kmag & \QNIR & \TE 
    }
    \bottomrule 
\end{tabular}
\end{table*}

\begin{table*}
\caption{Probable members of open cluster $ESFERA\,3$ from photometric observations}
\label{table:esfera3p}
\centering
\fontsize{9pt}{10pt}\selectfont
\setlength{\tabcolsep}{3pt}
\begin{tabular}{c c r c c c c c c l} 
    \toprule
    $GAIA-DR3$ & $GLON$ & $GLAT$ & $G$ & $G_{BP}-G_{RP}$ & $J$ & $H$ & $K_s$ & $Q_{NIR}$ & $SpT$ \\
    \midrule 
    5883...& $[{\circ}:':'']$ & $[{\circ}:':'']$ & $[mag]$ & $[mag]$ & $[mag]$ & $[mag]$ & $[mag]$& & \\ 
    \midrule 
    \begin{filecontents*}{Esfera3P.csv}
GAIA,GLON,GLAT,Gmag,BPRP,Jmag,Hmag,Kmag,QNIR,TE
194941656356480,321:56:08.3,-00:00:52.8,15.85,2.99,12.06,11.13,10.67,0.15,YSO II
214526707667072,321:59:49.4,-00:04:20.3,16.29,1.90,13.80,13.40,13.20,0.05,F$^-$
214973385226112,322:02:00.9,-00:03:46.3,15.29,2.35,12.06,11.48,11.19,0.10,B$^+$
215003418857472,322:03:00.7,-00:03:46.2,15.37,2.03,12.57,11.67,10.87,-0.47,YSO II
216210335393536,322:03:07.3,00:00:39.2,16.39,2.62,13.19,12.64,12.35,0.06,B$^+$
216622652247808,322:05:07.4,-00:02:00.5,16.78,1.76,14.58,14.05,12.82,--,F$^-$
216863170424576,322:07:37.0,-00:03:01.4,14.14,1.50,12.07,11.74,11.59,0.08,B$^+$
217172408108160,322:07:53.2,-00:00:21.3,14.90,1.28,13.11,12.83,12.71,0.09,A$^-$
217172408111232,322:08:03.6,-00:00:19.8,15.88,1.52,13.66,12.89,12.51,0.12,A$^+$
217412926228736,322:03:45.3,-00:01:00.7,16.74,2.44,13.70,13.11,12.82,0.09,A$^-$
218031401599744,322:08:35.0,00:00:53.6,15.88,1.81,13.54,13.03,12.88,0.24,A$^-$
219057869997952,322:00:21.5,00:02:04.4,17.63,2.10,15.04,14.21,13.71,-0.02,F$^+$
219440150823424,322:02:28.7,00:03:08.6,16.04,1.70,13.75,13.37,12.88,--,A$^+$
221020698860928,322:04:22.2,00:05:48.8,16.92,2.39,13.98,13.43,13.25,0.22,A$^-$
223151002640256,322:10:52.3,00:00:11.5,16.31,2.50,12.91,12.07,11.77,0.35,B$^+$
232221972981504,322:00:23.4,00:07:59.2,16.63,2.63,13.40,12.70,12.37,0.13,B$^+$
232359411945472,322:01:49.9,00:09:04.6,15.85,1.45,13.83,12.41,11.93,0.61,A$^+$
234524076184960,322:07:50.9,00:11:59.5,16.33,2.55,13.09,12.16,11.77,0.28,B$^+$
235344386772224,322:03:35.3,00:15:12.5,16.52,1.84,14.17,13.63,13.43,0.20,F$^-$
236826178920448,322:00:58.9,00:16:51.5,15.10,2.27,12.20,11.45,11.17,0.28,B$^+$
\end{filecontents*}
\csvreader[
        head to column names, 
        separator=comma,
        late after line=\\
    ]{Esfera3P.csv}{}{
        \GAIA & \GLON & \GLAT & \Gmag & \BPRP & \Jmag &  \Hmag & \Kmag & \QNIR & \TE 
    }
    \bottomrule 
\end{tabular}
\end{table*}

\begin{table*}
\caption{Probable members of open cluster $ESFERA\,4$ from photometric observations}
\label{table:esfera4p}
\centering
\fontsize{9pt}{10pt}\selectfont
\setlength{\tabcolsep}{3pt}
\begin{tabular}{c c c c c c c c r l} 
    \toprule
    $GAIA-DR3$ & $GLON$ & $GLAT$ & $G$ & $G_{BP}-G_{RP}$ & $J$ & $H$ & $K_s$ & $Q_{NIR}$ & $SpT$ \\
    \midrule 
    5886...& $[{\circ}:':'']$ & $[{\circ}:':'']$ & $[mag]$ & $[mag]$ & $[mag]$ & $[mag]$ & $[mag]$& & \\ 
    \midrule 
    \begin{filecontents*}{Esfera4P.csv}
GAIA,GLON,GLAT,Gmag,BPRP,Jmag,Hmag,Kmag,QNIR,TE
257425123101568,322:17:06.0,00:21:00.7,16.9,1.8,14.4,14.2,13.9,-0.2,G$^-$
348100474124800,322:28:00.7,00:34:00.6,17.2,1.8,15.0,14.5,14.2,0.0,G$^-$
349440504533248,322:30:08.3,00:59:00.7,16.1,1.6,14.0,12.0,11.6,1.3,F$^-$
350643094839552,322:42:00.4,00:25:00.7,16.0,1.5,14.0,13.6,13.5,0.2,F$^-$
351364649407360,322:20:13.0,00:38:00.8,13.6,1.1,12.0,11.8,11.7,0.0,B$^+$
351429059257984,322:20:16.6,00:46:00.8,12.0,1.1,10.6,10.4,10.3,-0.1,B$^+$
352429801298816,322:25:45.5,00:50:00.8,16.7,1.8,14.0,13.2,12.8,0.0,G$^-$
352567240269312,322:27:12.9,00:25:00.8,16.9,2.8,13.5,12.9,12.5,0.1,B$^+$
353220075304320,322:24:33.1,00:15:00.8,16.6,1.6,14.1,13.7,13.4,0.1,F$^+$
353769831159936,322:27:51.4,00:36:00.8,16.3,1.6,14.3,13.7,13.5,0.2,F$^-$
363287478678016,322:20:38.6,00:19:00.8,16.5,1.6,14.3,13.8,13.7,0.3,F$^+$
363386248420736,322:17:49.0,00:51:00.8,17.1,1.8,14.6,13.2,12.8,0.8,F$^+$
363871594262144,322:21:26.4,00:27:00.8,14.3,1.2,12.6,12.4,12.2,0.0,A$^-$
364318270911360,322:20:26.4,00:23:00.9,15.7,1.5,13.7,13.3,13.2,0.2,F$^-$
\end{filecontents*}
\csvreader[
        head to column names, 
        separator=comma,
        late after line=\\
    ]{Esfera4P.csv}{}{
        \GAIA & \GLON & \GLAT & \Gmag & \BPRP & \Jmag &  \Hmag & \Kmag & \QNIR & \TE 
    }
    \bottomrule 
\end{tabular}
\end{table*}

\begin{table*}
\caption{Probable members of open cluster $ESFERA\,5$ from photometric observations}
\label{table:esfera5p}
\centering
\fontsize{9pt}{10pt}\selectfont
\setlength{\tabcolsep}{3pt}
\begin{tabular}{c c c c c c c c r l} 
    \toprule
    $GAIA-DR3$ & $GLON$ & $GLAT$ & $G$ & $G_{BP}-G_{RP}$ & $J$ & $H$ & $K_s$ & $Q_{NIR}$ & $SpT$ \\
    \midrule 
    58838...& $[{\circ}:':'']$ & $[{\circ}:':'']$ & $[mag]$ & $[mag]$ & $[mag]$ & $[mag]$ & $[mag]$& & \\ 
    \midrule 
    \begin{filecontents*}{Esfera5P.csv}
GAIA,GLON,GLAT,Gmag,BPRP,Jmag,Hmag,Kmag,QNIR,TE
37086485049088,324:13:15.1,00:45:59.2,12.98,3.41,9.06,7.95,7.55,0.44,CLIII-I
37915429150592,324:09:25.2,00:50:30.0,17.23,1.94,14.77,14.29,14.03,0.02,F$^-$
38911861950720,324:13:58.5,00:41:59.1,16.28,1.92,13.87,13.32,13.09,0.16,BSS
40698568446208,324:18:05.0,00:47:48.3,16.95,2.86,13.72,13.12,12.88,0.19,CLIII
40732928203776,324:19:12.0,00:48:28.8,16.78,3.02,12.91,12.33,11.51,-0.80,CLIII
41832439453056,324:25:30.1,00:47:08.8,16.93,2.04,14.41,13.95,13.75,0.12,F$^-$
42210396958720,324:20:07.3,00:48:13.2,15.29,1.84,--,--,--,--,BSS
43992792751232,324:18:21.6,00:51:02.7,15.50,2.77,12.14,11.22,10.96,0.48,CLIII
44100182546944,324:14:25.5,00:51:07.6,16.39,2.49,13.42,12.94,12.64,-0.02,CLIII
44375059980800,324:14:18.1,00:52:55.5,16.06,2.67,12.96,12.48,12.18,-0.02,CLIII
44409419719040,324:13:21.5,00:53:00.8,16.28,1.82,13.93,13.50,13.42,0.28,BSS
44512499441280,324:15:29.4,00:53:20.2,13.12,1.58,11.09,10.77,10.54,-0.07,BSS
44649938394624,324:17:25.1,00:51:41.5,14.58,1.59,12.53,12.21,12.01,-0.01,BSS
44993535795840,324:16:38.2,00:54:21.6,15.34,2.54,12.19,11.31,10.99,0.34,CLIII
45165334464256,324:19:28.6,00:49:12.4,16.35,2.25,13.64,13.17,12.94,0.09,CLIII
91825860877184,324:14:10.8,00:57:19.5,15.39,1.76,13.12,12.79,12.55,-0.07,BSS
\end{filecontents*}
\csvreader[
        head to column names, 
        separator=comma,
        late after line=\\
    ]{Esfera5P.csv}{}{
        \GAIA & \GLON & \GLAT & \Gmag & \BPRP & \Jmag &  \Hmag & \Kmag & \QNIR & \TE 
    }
    \bottomrule 
\end{tabular}
\end{table*}

\end{appendix}
\end{document}